\documentclass[a4paper,11pt]{article}
\pdfoutput=1 % if your are submitting a pdflatex (i.e. if you have
\usepackage{jheppub}  %for details on the use of the package, please
\usepackage{simpler-wick}
\usepackage[T1]{fontenc} % if needed
\usepackage{physics}
\usepackage{graphicx}

\def\a{\alpha}

\def\beq{\begin{equation}}
\def\eeq{\end{equation}}
\def\bea{\begin{eqnarray}}
\def\eea{\end{eqnarray}}

\def\bit{\begin{itemize}}
\def\eit{\end{itemize}}

\def\c{\chi}

\def\baa{\begin{array}}
\def\eaa{\end{array}}

\def\simgt{\mathrel{\lower2.5pt\vbox{\lineskip=0pt\baselineskip=0pt
           \hbox{$>$}\hbox{$\sim$}}}}
\def\simlt{\mathrel{\lower2.5pt\vbox{\lineskip=0pt\baselineskip=0pt
           \hbox{$<$}\hbox{$\sim$}}}}

\def\bfc{\begin{figure}\begin{center}}
\def\efc{\end{center}\end{figure}}

\title{\boldmath Probing the initial state of inflation: analytical structure of cosmological correlators}

\author[a,1]{Diptimoy Ghosh}
\author[a,b,2]{Amartya Harsh Singh}
\author[a,3]{Farman Ullah}
\affiliation[a]{Indian Institute of Science Education and Research, Pune, India}
\affiliation[b]{SISSA International School for Advanced Studies, Via Bonomea 265, 34136, Trieste, Italy}

\emailAdd{diptimoy.ghosh@iiserpune.ac.in}
\emailAdd{amartya.harshsingh@students.iiserpune.ac.in}
\emailAdd{farman.ullah@students.iiserpune.ac.in}
\abstract{We study the analytic structure of in-in correlation functions in a deSitter background. The aim of this study is to probe the initial conditions for inflation through the features of correlation functions of the  field fluctuations, and understand precisely how an in-in correlator responds to particles in the initial state. We emphasize that the choice of vacuua and the corresponding particle interpretation for these fluctuations is flexible, and we clarify the role of this choice at the level of calculations and their diagrammatic interpretation. We consider several possibilities aside from the standard Bunch Davies (BD) vacuum prescription for the initial state, and trace the change in pole structure as one begins adding excitations; starting from just a single particle, to highly excited states and special cases such as a coherent state. We also go beyond the pole structure of the bispectrum, and calculate the 4 point classical and quantum correlators.
We illustrate - with the example of coherent states - the subtleties in concluding a Bunch Davies initial state from the absence of physical poles in the bispectrum, which is interesting in light of some recent literature.
Initial states with a finite number of excitations are plagued with disconnected diagrams isolated in phase space, and we highlight their implications on the observation of these signals, and how the situation changes as one begins to excite more and more particles. We also comment about the implications of various initial conditions on the squeezed limit of the bispectrum.
These new pole structures are a direct consequence of mixing of positive and negative frequency modes which is a characteristic of curved spacetimes; in particular, we see in detail how particles in an initial state replicate mode mixing structures. This study aims to clarify the missing details that link quantum and classical initial conditions, and sharpen our understanding of in-in correlators in inflation.}

\begin{document} 
\maketitle
\section{Introduction}
It is now widely believed that the structure in our universe was seeded by quantum mechanical fluctuations generated during an epoch of exponential expansion known as Inflation\cite{Guth:1980zm,Achucarro:2022qrl}. During this period, these quantum fluctuations were stretched to super-horizon distances with unchanging amplitudes\cite{Mukhanov:1981xt} which later gave rise to the rich structure that we observe around us. The predictions of inflationary models are consistent with observations but there remain some unanswered questions, such as what were the initial conditions for inflation? In most models, the perturbations start their life in a special kind of vacuum state known as the Bunch-Davies vacuum but this assumption is not free of contention(\cite{Agullo:2010ws,Ganc:2011dy, Holman:2007na, Easther:2002xe, Brandenberger:2002hs, Meerburg:2009ys}). The primary motivation for this is that we know nothing about the (possibly Transplanckian) physics that preceded inflation, and assuming any boundary condition in the far past is an extrapolation of intuition to scales we don't understand. It turns out that the most general vacuum state possible is the general Bogolyubov transformed Bunch-Davies state (\cite{Holman:2007na, Ganc:2011dy}) and the most general vacuum state invariant under de Sitter isometries is the so called $\alpha$ vacuum\cite{Shukla:2016bnu,PhysRevD.32.3136}. The power spectrum alone cannot distinguish between these different choices of vacua and initial states. Therefore, we turn to higher point functions e.g., Bispectrum\cite{Acquaviva_2003,Maldacena_2003,Weinberg_2005,Bartolo_2004} and their analytic structures to look for distinguishing features. It is believed that perturbations that had a classical origin can give rise to present observations at the level of the power spectrum, without appealing to any quantum origin (\cite{Green:2020whw, Green:2022fwg}); see \cite{Agullo:2022ttg} for interesting recent developments. The definition of what pole structure one calls classical, however, is unclear and it is among the purposes of this paper to clarify this point. The crucial insight is to study the analytic structure of the correlation functions as a function of the initial conditions, and we find that a simple definition of classical - such as a state with a large number of particles - is not enough to indicate the presence of a different pole structure than that of a true quantum vacuum. Coherent states for a Bunch Davies observer which are clearly excited - under certain assumptions - behave exactly like a Bunch Davies vacuum at the level of correlators. Moreover, any other choice of initial state aside from the Bunch Davies vacuum and it's coherent states displays a recurring presence of physical poles. Therefore, in a sense, the analytic structure arising from a Bunch Davies vacuum or coherent state is very unique; with everything else showing a common additional structure in the form of \textit{physical poles}. It is the purpose of this work to trace precisely the origin of this difference, and study in detail the new features that in-in correlators start displaying as one changes the nature of the initial state. 

%Indeed, if one takes seriously the correspondence between dS and a 3D Euclidean CFT, then imposing OPE consistency on the momentum-space correlators(Fourier transformed from the easily fixed position space versions) singles out total-energy poles that are characteristic to the BD vacuum as the choice of initial state. The complete Fourier transform contains additional physical poles that OPE consistency kills, but if we depart from this correspondence then these poles survive. This shouldn't be a surprise, since it's long known that, for example, the alpha-vacuum does not correspond to any exact CFT, but one that is marginally deformed. Therefore imposing OPE on it's correlators is incorrect, and one would have all possible pole structures, as we explicitly report(as to \cite{Holman:2007na, Ganc:2011dy}, etc). It must be emphasized that demanding OPE isn't automatic-there is no clear dS/CFT correspondence, and what one is really doing is using the de Sitter isometries-which happen to be those of a Euclidean CFT-to constrain correlators. Generically the pole structure is diverse, and therefore any evidence in favor of Bunch Davies as the initial state with its characteristic pole structure will be insightful about the truth of a dS/CFT correspondence.

The different pole structures can ultimately be tied down to the presence or absence of mixing of positive and negative frequency modes-in curved space, these  can naturally arise depending on the initial condition. When we think of inflation as being driven by some Effective field theory, we need to understand what features of the physics before inflation are important to understand. New physics can manifest itself as either irrelevant operators(consistent with slow-roll conditions) or as initial state modifications (backreaction effects must be consistently included/dropped - this is a subtle issue that we don't explore here). At the level of the power spectrum, given a cutoff $\Lambda$; initial state effects can add corrections of the order $\frac{H}{\Lambda}$, much larger than the $\frac{H^2}{\Lambda^2}$ effects coming from irrelevant operators \cite{Kaloper:2002cs}, and therefore initial state effects cannot be disregarded, despite their somewhat more obscure origin(see \cite{Cabass:2022avo} for a recent review of the EFT of cosmology). The observed power spectrum then allows us to infer bounds on the scale of new physics, $\Lambda$. More generally, they constrain the number density of particles in the initial state by constraining the Bogolyubov parameters, that, in this case, modify also the normalization of the power spectrum. However, higher point correlation functions are more sensitive probes(\cite{Holman:2007na}) for the simple reason that they are more sensitive to interactions of particles in the initial state, and have always been an object of interest either from a bootstrapping perspective or as a tool to systematically trace signatures of new physics in these correlators \cite{https://doi.org/10.48550/arxiv.2205.10340,Pajer:2020wxk, Baumann:2015nta,Arkani-Hamed:2018kmz, Arkani-Hamed:2015bza, Pimentel:2022fsc, Chen:2022vzh, Baumann:2022jpr,Flauger:2022hie}; see \cite{Benincasa:2022gtd} for a recent view. For this matter, we turn our attention to the bispectrum and the 4-point function \cite{Seery_2007,Chen_2009,Senatore_2011, Chen:2006nt,Bonifacio:2021azc} with various initial conditions, and observe how their structure changes. We  don't make any pretence about the kind of physics before inflation that leads to the choice of initial state. In particular, we emphasize that the universe has no obligation to exist in the vacuum state of the perturbations.

 Correlation functions of the inflaton (more precisely, the scalar mode of the ADM metric after removing gauge freedom) are an obvious tool to study the physics of inflation, encoding both the nature of the interactions in the EFT of inflation, as well as the initial conditions for inflation itself. These correlators, on the other hand, must not be sensitive to whatever fundamental physics drove inflation.We have to be careful about the nature of singularities that are generated by different interactions, as well as the need for an IR regulator. For instance, the difference between a $\zeta^3$ and a $\zeta'^3$ interaction in an in-in calculation reflects in the singularities they generate in a correlation function. The former generates logarithmic singularities, whereas the latter generates simple poles. The log divergences are contact terms in Fourier space, and their origin can be traced to the fact that conformal symmetry constrains position space correlators such that they are singular at coincident points. This relationship deserves a detailed analysis of it's own(\cite{Kehagias:2012pd,Antoniadis:2011ib}), and we don't dwell on this subject in this work. We'll instead work with a derivative interaction where the poles are simple poles in momenta and no regulator is required, and we see how the correlation functions with this specific interaction respond to changes in initial conditions. Three point functions, in particular, are especially interesting because they are a measure of the `non-gaussianity', which is a direct measure of the existence of interactions in the inflationary action. The important thing to notice is the word 'interaction', and this is where the nature of deSitter geometry and corresponding vacuua becomes important. In flat space, one typically has the convenience to turn off all interactions in the far past and future, so that the vacuum state at the asymptotic boundaries is free of excitations. In curved backgrounds, in contrast, one doesn't have this luxury, and in fact, the notion of particles and excitations is not well defined. This has consequences on the nature of correlation functions evaluated with different initial conditions. For example, if the initial state is excited relative to, say, the Bunch Davies observer, then it is possible in some cases to find an observer which sees no particles. These cases are precisely those which are related via a Bugolyubov transform. The interpretation of correlators as Feynman diagrams is therefore very different - the diagrams for a Bunch Davies observer contains disconnected pieces because there are more particles in the initial state than those that can interact at a vertex-these will then have support only at localised points in phase space. The diagrams for an alpha-observer however are fully connected as only the field insertions interact at the vertex.
 
 In either case, we see the presence of physical poles where the singularities can be reached in the physical region. It is tempting to relate this to an interpretation in terms of particle scattering and decays in the initial state, but as we remarked, this picture is observer dependent.  Clearly, the situation is more complicated than a relative particle density between the vacuum state of the modes and the initial state for inflation. An initial state which is an alpha vacuum at the start of inflation produce a similar structure to the one mentioned above, but here it is clear that there is no particle self interaction interpretation that can be ascribed. If one insists to work with an observer who sees particles, then it is true that an analogy can be drawn with flat space scattering processes, but one must then deal with the fact that the phase space where these poles have support is restricted.  Therefore, it isn't straightforward to trace various pole structures to the absence/presence of particles in the initial state. One has the same problem in flat space-LSZ only gives us information about fully connected diagrams that have the right pole structure, but the S-matrix contains contributions from both connected and disconnected diagrams in a highly excited state.
 
 For example, if we had a cubic vertex $\phi^3$ and we computed the leading order contribution to 
 
 \begin{equation}
    \int d^4x\langle n_{out}|\phi^3(x)|n_{in}\rangle\supset\int d^4x\langle n_{k_1}+1,n_{k_2}+1,n_{k_3}-1,..|n_{in}\rangle \sim \delta(k_1-k_2-k_3)A_{1\to 2}
 \end{equation}
 
 where $n\equiv\otimes_{k}(a^\dagger_k)^n|0\rangle/\sqrt{n!}$. Here, $A_{1,2}$ is now a disconnected amplitude with $k_1,k_2,k_3$ interacting at a vertex, while the rest of  the $n_k$'s spectate. LSZ kills this diagram, and therefore it doesn't contribute to the fully connected part of the S-matrix; it exists nonetheless and since with inflation we don't have the luxury of an S-matrix formalism, we must make our peace with such processes.

\begin{figure}
     \centering
     \includegraphics{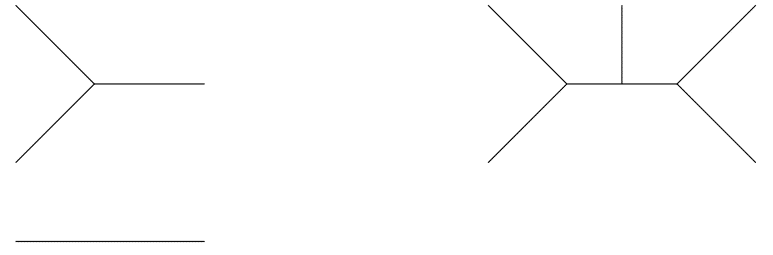}
     \caption{(Left) Disconnected diagram at order  $\lambda$. (Right) Fully connected at order $\lambda^3$}
     \label{fig:flat}
 \end{figure}

 We are typically interested only in the fully connected part of the S-matrix, but such a requirement cannot be imposed when the initial states are highly occupied. Going to higher orders in perturbation theory partially remedies this situation. For instance, consider a $3\to 2$ process in flat space with a cubic vertex. The Feynman diagrams for the first two orders in perturbation theory are shown in figure \ref{fig:flat}. We see that going to higher orders in perturbation theory is one way to generate connected diagrams. However, this problem only worsens as we include more and more particles in the external states, because then the fully connected diagrams get harder and harder to measure. In curved space in-in correlators, as we shall see, the recipe to circumvent these is to switch to a frame where there are no particles and the state is the vacuum with respect to the new modes. We will then be left with only fully connected diagrams, whose pole structure can be analysed. Generic excited states display physical poles, but their signatures are clouded by the restricted points in phase space where they have support. We will see that these physical poles occur when particles in the initial state interact at a vertex along with some of the field insertions, while the remaining field insertion is a spectator and therefore has support only when it's momentum coincides with one of the momenta of the spectator particles in the initial state. If we have all possible momenta in such a state(what one would expect in a 'classical' state), then atleast one of these physical poles will be observable, but at an isolated point in phase space for a given momentum. Scanning all possible momenta for the field insertions will then generate a measurable physical pole. In the case where the initial state is a Bugolyubov transform of the Bunch Davies vacuum, the physical poles are seen transparently. 
 
 Interestingly, there is a class of excited states over the Bunch Davies vacuum-the so called coherent state $|C\rangle$ \cite{Kundu_2012} 
 
 \begin{equation}
     a_{\vec{k}}|C\rangle=C(\vec{k})|C\rangle
 \end{equation}
 
 which, under certain assumptions, reproduces the pole structure for the vacuum state i.e. it displays no physical poles despite being an excited state. This is in direct contradiction with the supposed association of physical poles with excited states. If instead we had a coherent state that was built over not the Bunch Davies vacuum, but that of its Bugolyubov transform, then physical poles are restored. 
 
 Let us elaborate on the role of disconnected pieces. Once the order in perturbation theory is fixed, we must allow the possibility of other particles to spectate the interaction and contribute as disconnected pieces. LSZ will have nothing to say about this, and therefore the flat space intuition for physical poles in inflation is premature. This manuscript is devoted to studying in detail the origin of pole structures in in-in correlators and the information it carries about the initial state of inflation, and probing the relationship between a highly excited quantum state versus a 'classical' state.
 
 The broad outline of this manuscript is as follows. In section \ref{sec:calc}, we will analyse in-in 3-point correlators for a variety of initial states, including the anomalous case of coherent states. In addition, subsection \ref{sec:fourpt} is devoted to checking explicitly, by computing the 4-point classical and quantum in-in correlation function, that going to higher orders in perturbation theory doesn't change the basic inferences we draw from the bispectrum about the singularity structure, extending the result of \cite{Green:2020whw}. In section \ref{sec:LSZ}, we review the features of LSZ in flat space, and the corresponding intuition to be drawn for in-in correlators. Next, in \ref{sec:interactions} we study the role of derivative interactions that require no IR regulator in generating singularities that are simple poles.   In section \ref{sec:Maldacena} we discuss the squeezed limit of the bispectrum. Finally, in 
 section \ref{sec:classical} we clarify the link between physical poles in classical correlators versus those in quantum in-in correlators.

 \section{From flat space to curved space}

%1) General background, mode expansion, Bogulyibov transformations

%2) Comments on structure of vacuum, Bunch Davies vacuum

%3) Things  specific to deSitter, bulk and boundary propagators, conformal boundary

%4) In-in formalism, relation between bulk propagators and feynman propagators and correlators on a constant time hypersurfae

%5) Note on duality, analytic continuation to AdS (Anninos)

\subsection{Background}\label{sec:background}

Provided with the field content, background geometry, and symmetries of the system; one can write an action functional that describes the theory. The equations of motion following from this action are solved in terms of \textit{mode functions}  $u_k(t,\vec{x})$ that multiply some coefficients $a_k$, and their hermitian conjugates. In time independent, flat backgrounds we have plane wave solutions for the mode functions of free fields; $u_k(t,\vec{x})=\frac{1}{\sqrt{2\omega_k}}e^{i\vec{k}{x}-\omega t}$. We promote $a_k, a_k^\dagger$ to operators that satisfy canonical commutation rules that then get imparted to the fields themselves, and we say we have canonically quantized the theory. Studying this theory then amounts to studying the corresponding Hilbert space, and this discussion begins by first identifying a \textit{vacuum state} from which other states are realised, defined as $a_k|0\rangle=0$. In flat space, it is uniquely fixed by demanding invariance under the underlying Poincare group. This means that the mode functions and the quantization prescription of the theory is unique. This is traced to the fact that $u(k)$ is Lorentz invariant, and this is where the situation drastically changes in more general backgrounds. 

If we allow for diffeomorphism invariance-a much larger group-then the $x$ and $t$ dependence of the mode functions can change since ($\vec{k}\cdot \vec{x}-\omega t$) need not be invariant anymore. In fact, plane waves simply needn't be solutions to equations of motion in general backgrounds. Moreover, in time dependent backgrounds-as is the case  with inflation-$\omega$ has a time dependence, which reflects as an additional time dependence in the mode functions.  Therefore, mode functions are in general not unique. Being solutions to second order equations of motion, they require two additional inputs to be fixed. Consequently, different mode expansions are distinguishable in the sense that they differ, for example, in their measurement of particle density in a given state. Note that this difference lies in their different basis of modes (and consequently creation/annihilation operators), but the definition of a field itself ofcourse is invariant under these transformations. Again, one has to keep in mind that this phenomenon occurs simply by allowing accelerated coordinate frames, even in flat space, and are therefore tied to the definition of an inertial observer. In special relativity these are restricted to those obtained from Poincare transformations on a Minkowsky background; in general relativity we allow for the more general case of whatever isometry preserves the background geometry.

Unlike flat-space where demanding the vacuum to be invariant under the isometries of the background metric uniquely fixes the vacuum state, in de Sitter the invariance under isometries does not completely fix the vacuum state\cite{PhysRevD.32.3136}. Therefore, there is a class of vacua known as alpha vacua which are invariant under de Sitter isometries and parametrized by $\alpha$\cite{PhysRevD.32.3136,Shukla:2016bnu}. The mode function corresonding to $\alpha$ vacua are given by 
\begin{align}
\label{alpha}
    u_{k}(\tau)=\frac{\Delta_{\zeta}}{\sqrt{k^{3}}}\left(\alpha(1-ik\tau)e^{ik\tau}+\beta(1+ik\tau)e^{-ik\tau}\right)
    \end{align}
where,
\begin{equation}
    |\alpha|^{2}-|\beta|^{2}=1
\end{equation}
 \eqref{alpha} is the solution of Mukhanov-Sassaki equation
 \begin{equation}
     \zeta_k''-\frac{2}{\tau}\zeta_k'+k^{2}\zeta_k=0
 \end{equation}

If we call the $\alpha$ modes $v_k$ and the Bunch Davies modes by $u_k=(1-ik\tau)e^{ik\tau}$, then we can compare the associated initial conditions through
 \begin{align}
  \dot{v_k}(\tau_0)=-i\omega'_kv_k(\tau), \dot{u}_k(\tau_0)=-i\omega_ku_k(\tau)\implies \frac{\alpha_k}{\beta_k}=\frac{\omega'_k-\omega_k}{\omega'_k+\omega_k}
\end{align}

 The Bunch Davies vacuum ($\alpha=1$, $\beta=0$) corresponds to one particular choice of $\alpha$ vacua. This choice of vacuum is usually considered the most natural one since the modes that we are interested in were far inside the horizon at the beginning of inflation ($\tau\rightarrow -\infty$ limit ) therefore these modes do not "feel" any de-Sitter curvature and can therefore can be treated as Minkowski modes. But as mentioned above this is not a settled issue. It is important to note that no co-moving/free falling particle detector will register any particles in a  particular $\alpha$ vacuum since all these observers are related by de Sitter isometries and these vacua are invariant under them. \par

 \subsection{What we learn}

The purpose and findings of this present work can be summarized as follows-
\begin{itemize}

\item Is there a unique pole structure for all in-in correlators? What distinguishes the physics leading to different singularities? \textit{No, the pole's functional form depends on the choice of interaction. The physical origin we aim to see lies not in the functional form of the singularity but the momentum configuration for which it occurs.} See \ref{sec:interactions} for a discussion.
    
    \item Are total energy poles a characteristic of only the Bunch Davies vacuum intial state with no excitations? Do physical poles arise only if the initial state was an excited state? \textit{What is true is that most excited states relative to a Bunch Davies vacuum display physical poles. However, a class of coherent states displays the same pole structure as a state with no Bunch Davies particles, and this is an important anomaly that we wish to highlight}. This result is derived in \ref{eq:coherent}.

    \item Do initial states with a fixed number of Bunch Davies particles display observable physical poles?Is it necessary for each mode to have a high occupation number in the initial state to replicate the structure of a classical correlator? \textit{With a finite number of particles, the physical poles are isolated in phase space. One needs to excite all possible momenta to guarantee an observable physical pole. The important ingredient is to not miss out on regions in phase space, and to excite every mode atleast once. This state however has a large number of particles anyway, as one would expect for a classical state} See \ref{oneparticleBD} and eq. \ref{eq:disc0},\ref{eq:disc1},\ref{eq:disc2} to understand the support for these new poles.
    
    \item If the initial state is a vacuum state, then do we not see physical poles? \textit{The notion of vacuum state is not unique, and for any other vacuum other than Bunch Davies, one sees physical poles despite there being an observer who sees no particles in that state.} See \ref{sec:alpha}for discussion. 
    \item What is the underlying reason for similar mathematical structures of classical correlators and those from excited quantum states? \textit{The classical Green's function has both signs of modes. If an excited quantum state is the vacuum for an observer related via a Bogolyubov transform to the Bunch Davies vacuum, then in the picture where we work in the frame of this observer who sees a vacuum, the associated Wightman functions have both signs of modes too. If one instead continues to use the Bunch Davies picture, then the physical poles arise because of initial state particles interacting at a vertex with the field insertions produce modes with a relative sign between momenta.} See \ref{sec:classical} for details.

    %{\color{green} \item Does OPE consistency single out Bunch Davies as a special initial state?\textit{Yes, OPE singles out the case of BD boundary conditions for both the modes and the initial state}
    %\item Are all these classes of correlators physically sensible quantities? \textit{yes}}

    \end{itemize}

\section{In-In Correlation functions and associated analytic structures}\label{sec:calc}

In this section, we will report case by case the analytic structure for correlators corresponding to various choices of initial states and mode functions(i.e. the choice of vacuum). We use the In-In formalism (see Appendix \ref{in-in}) to compute quantum correlators. While our primary interest will be the 3 point function, we also write down the power spectrum for completeness. We will also look at the structure obtained with a coherent initial state and contrast it with a general excited initial state. Finally, we will look at the 4 point function calculated using both the in-in formalism and a classical perturbation theory formalism, and check if our intuition after studying 3 point functions holds true there too.

The results in this section will be organized in the following manner. We will first fix the mode functions i.e. the vacuum state of the perturbations to be either BD or alpha-modes, and then compute the correlation functions with the following choices of initial states-

\begin{itemize}
    \item The initial state coincides with the vacuum state
    
    \item The initial state is a one particle state of definite momentum
    
    \item The initial state is an excited state with all modes occupied
    
    \item The initial state is a highly excited state which is a  \textit{Bogolyuobov transform of Bunch Davies}.
    
    \item The initial state is a coherent state for either a Bunch Davies or a Bogolyubov transformed observer
\end{itemize}

The results, to be detailed in the following sections, are summarized in the following table-

\begin{center}
     \begin{tabular}{c|c}
         Initial state & Pole structure  \\
         \hline
          BD vacuum & $k_t$ pole\\
          Excited over BD with finite modes present & disconnected physical poles, connected $k_t$ pole\\
          Excited over BD with all modes present & connected $k_t$ pole, observable physical pole\\
          Excited over BD but a BG & both physical and $k_t$ poles\\
          Excited over BD but coherent & $k_t$ pole\\
          Excited over BD but coherent for some BG & both $k_t$ and physical poles\\
          Classical & both $k_t$ and physical poles
     \end{tabular}
 \end{center}\label{table}

\subsection{Normalization of the expectation value}

We must recognize that computing the expectation value for any state requires that we divide also by the norm of that state, which needn't always be one. For example, with a $|p\rangle$ state the norm is proportional to $\delta^{(3)}(\vec{p}-\vec{p})$ which is formally divergent. However, a correlation function of the form $\langle p|\zeta_1\zeta_2\zeta_3|p\rangle$ will also contain factors of $\langle p|p\rangle$ when we contract the two initial states together(nothing prevents us from doing it at the level of correlation functions, this is just a disconnected diagram). Therefore, after dividing by the normalization, we will see that this disconnected piece's contribution vanishes and we're left with a finite term whose pole structure can be  read off. All other terms will be suppressed by this (infinite) normalization factor, except when they have localised  support in phase space through other delta function singularities, in which case they can peak at \textit{those} points in phase space, but those points only. If we take these correlators to represent NG, then these isolated pole structures(and therefore the corresponding initial states) are ruled out by observation. It is important to note that $\delta^{(3)}(0)$, wherever it occurs, will be regulated by a volume $V$, which we will see ultimately drops out of the correlator. Also, the factor of $2E_p$ that relativistically normalize a $|p\rangle$ drop out. In the below correlators, therefore, such a normalization is always implicit, even if not explicitly written.

\subsection{BD vacuum initial state}

We review the results from the most common choice of initial state.

\subsubsection{Two point function}At $O(\lambda^{0})$ the two-point function is given by
\begin{equation}
    \bra{0}\zeta_{k_1}\zeta_{k_2}\ket{0}=\frac{\Delta^{2}_{\zeta}}{k_1^{3}}\delta^{3}(\vec{k_1}+\vec{k_2})
\end{equation}
\subsubsection{Three point function}
To calculate
\begin{align}
     \bra{0}\zeta_{k_1}\zeta_{k_2}\zeta_{k_3}\ket{0}
     \end{align}
     At $O(\lambda^{0})$ the three point function vanishes so we compute at $O(\lambda)$\cite{Green:2020whw}
     \begin{align} 
        \bra{0}\zeta_{k_1}\zeta_{k_2}\zeta_{k_3}\ket{0}
=\frac{4H^{-1}\Delta^{6}_{\zeta}\lambda}{(k_1+k_2+k_3)^{3}k_1k_2k_3}\delta^{3}(\sum \vec{k_{i}})\label{eq:3pt}
     \end{align}

     \subsection{Excited initial state from Bogolyubov transform:$\alpha$-vacuum}\label{sec:alpha}
   
   In this section, we consider initial states which are excited relative to a Bunch Davies observer, but can \textit{also} be reached by a Bogolyuobov transform of the Bunch Davies vacuum. That is, we consider excitations that a different observer would see when observing a Bunch Davies vacuum. What this means is that there exists an observer for which such a state is actually the vacuum, while it has particles for a Bunch Davies observer. From a calculation standpoint, this is a massive simplification, because vacuum states are simpler to deal with than excited states. For concreteness, let us study an initial state that would correspond to an $\alpha$ vacuum, which is also de Sitter invariant and can be reached from the Bunch Davies initial state by a simple Bogolyuobov transform. 
   
   Given the Bunch  Davies mode functions $u_k(\tau)$ and it's set of creation and annihilation operators $a_k,a^\dagger_k$, the Bogolyuobov creation annihilation operators are $b_k=\beta^*a_k-\alpha^*a^\dagger_{-k}$
   
      where the $\alpha,\beta$ are parametrized by a single parameter(it's part of the 2 parameter family of dS invariant vacua, and we set one of the parameters to zero) $\tilde{\alpha}$; ${\beta,\alpha}={\cosh(\tilde{\alpha}),-i\sinh(\tilde{\alpha})}$. The $\alpha$ vacuum can be expressed in terms of the Bunch Davies expression through
   (\cite{Shukla:2016bnu})
   
   \begin{equation}
       |\alpha\rangle=\frac{1}{N}\exp\bigg(\frac{\alpha^*}{2\beta^*}\int \frac{d^3k}{(2\pi)^3}a^\dagger_ka^\dagger_{-k}\bigg)|0\rangle
   \end{equation}
   
  Where $N$ is an overall normalization constant. We see straightaway that it contains Bunch Davies excitations, but they come in pairs of equal and opposite momenta and therefore the state contains no net momentum. One would therefore hope that any disconnected pieces in the three point function will only have support at zero momentum for some of the external momenta(since there is no nonzero spectator particle momentum that requires to be conserved along with them)-an uninteresting case. Let us explicitly see how the calculation of the three point function goes through. We are interested in the object
  
  \begin{multline}
      \langle\alpha|\zeta_{k_1}\zeta_{k_2}\zeta_{k_3}|\alpha\rangle=\frac{1}{N^2}\bigg(\langle0|\zeta_{k_1}\zeta_{k_2}\zeta_{k_3}|0\rangle+\int \frac{d^3p}{(2\pi)^3}\frac{\alpha^*}{2\beta^*}\langle0|\zeta_{k_1}\zeta_{k_2}\zeta_{k_3}a^\dagger_p a^\dagger_{-p}|0\rangle\\
      \\+\int \frac{d^3p}{(2\pi)^3}\frac{\alpha}{2\beta}\langle0|a_{-p}a_p\zeta_{k_1}\zeta_{k_2}\zeta_{k_3}|0\rangle+\frac{|\alpha|^2}{4|\beta|^2}\int\frac{d^3p}{(2\pi)^3}\frac{d^3q}{(2\pi)^3}\langle0|a_{-p}a_p\zeta_{k_1}\zeta_{k_2}\zeta_{k_3}a^\dagger_q a^\dagger_{-q}|0\rangle +...\bigg)
  \end{multline}
   
   Let us name these terms $I,II,III,IV$ so that the correlator on the left is their sum. The terms can be arranged in powers of $\frac{\alpha}{\beta}$, and we therefore consider represenative examples from each power. To start with, we see that at zeroth power i.e. the term $I$, the bispectrum is the same as that for Bunch Davies vacuum, and we have a total $k$ pole only with no disconnected subgraphs. Let us move on to terms linear in $\alpha/\beta$; consider $III$ at first order in perturbation theory-
   
   \begin{multline}
       \int \frac{d^3P}{(2\pi)^3}\langle0|a_{-P}a_P\zeta_{k_1}\zeta_{k_2}\zeta_{k_3}|0\rangle=2\textrm{Im}\prod_{i=1}^3\int \frac{d^3p_i}{(2\pi)^3}e^{i\vec{p}_i\cdot\vec{x}}\int\frac{d\tau}{\tau}\int\frac{d^3P}{(2\pi)^3}\\
       \\\times \langle0|a_{-P}a_P\zeta_{k_1}\zeta_{k_2}\zeta_{k_3}\zeta'_{p_1}(\tau)\zeta'_{p_2}(\tau)\zeta'_{p_3}(\tau)|0\rangle
   \end{multline}

   \begin{multline}
      =2\textrm{Im}\prod_{i=1}^3\int \frac{d^3p_i}{(2\pi)^3}e^{i\vec{p}_i\cdot\vec{x}}\int\frac{d\tau}{\tau}\int\frac{d^3P}{(2\pi)^3}\langle a_P\zeta_{k_1}\rangle\langle a_{-P}\zeta'_{p_1}\rangle \langle \zeta_{k_2}\zeta'_{p_2}\rangle\langle \zeta_{k_3}\zeta'_{p_3}\rangle+\textrm{perm.}\\
       \\=\int \delta(\vec{P}+\vec{k_1})\delta(\vec{P}-\vec{p_1})\delta(\vec{k_2}+\vec{p_2})\delta(\vec{k_3}+\vec{p_3})(u^*_{k_1}(0)u'^*_{p_1}(\tau)) (u_{k_2}(0)u'^*_{p_2}(\tau))(u_{k_3}(0)u^*_{p_3}(\tau))\\
       \\=\delta(\vec{k}_t)2\textrm{Im}\int \frac{d\tau}{\tau}\prod_{i=1}^3\big(u_{k_i}(0)u'^*_{k_i}(\tau)\big)
   \end{multline}
   
   Thus we find, upto permutations(we have used that $u_{k_1}(0)$ is purely real in the first paranthesis)
   
   \begin{equation}
       III=\frac{H^6}{8k_1k_2k_3}2\textrm{Im}\int_{-\infty}^0 d\tau \tau^2e^{i(k_1+k_2+k_3)\tau}=\frac{H^6}{2k_1k_2k_3}\frac{1}{(k_1+k_2+k_3)^3}
   \end{equation}
   
\begin{figure}
       \centering
       \includegraphics{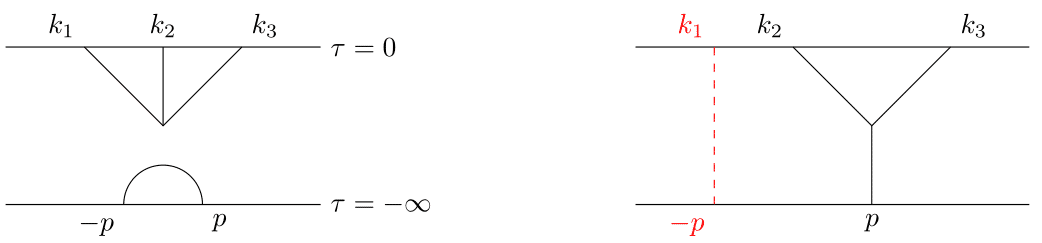}
       \caption{(Left) the initial state particles are spectators. (Right) the initial state particles interact at a vertex. There is a sum over all possible $p$ values.}
       \label{fig:2}
   \end{figure}

   We see that we only have a total energy pole. Figure \ref{fig:2} explains diagrammatically the situation.
   
   Let's check if this changes at the next order in $(\alpha/\beta)$. Take $IV$-
   
   \begin{multline}
              \int \frac{d^3P}{(2\pi)^3}\frac{d^3Q}{(2\pi)^3}\langle0|a_{-P}a_P\zeta_{k_1}\zeta_{k_2}\zeta_{k_3}a^\dagger_Qa^\dagger_{-Q}|0\rangle=2\textrm{Im}\prod_{i=1}^3\int \frac{d^3p_i}{(2\pi)^3}e^{i\vec{p}_i\cdot\vec{x}}\int\frac{d\tau}{\tau}\int\frac{d^3P}{(2\pi)^3}\\
       \\\times \langle0|a_{-P}a_P\zeta_{k_1}\zeta_{k_2}\zeta_{k_3}\zeta'_{p_1}(\tau)\zeta'_{p_2}(\tau)\zeta'_{p_3}(\tau)a^\dagger_Q a^\dagger_{-Q}|0\rangle
   \end{multline}
   
   This will obviously include disconnected pieces because there are more terms to contract than there are external points. For example, consider the contraction 
   \begin{equation}
   IV\supset \int \langle a_P\zeta_{k_1}\rangle\langle a_{-P}\zeta_{k_2}\rangle\langle \zeta'_{p_1}a^\dagger_Q\rangle\langle \zeta'_{p_2}a^\dagger_{-Q}\rangle\langle \zeta_{k_3}\zeta'_{p_3}\rangle\propto \delta({\vec{k}_3})\delta(\vec{k}_2+\vec{k}_1)
   \end{equation}
   
   which is disconnected and has support only when one of the momenta is zero. 
   
   Consider instead the contraction $$\langle a_P\zeta_{k_1}\rangle\langle\zeta_{k_2}\zeta'_{p_2}\rangle\langle\zeta_{k_3}\zeta'_{p_3}\rangle\langle \zeta'_{p_1}a^{\dagger}_{-Q}\rangle\langle a_{-P} a^\dagger_Q\rangle$$ This yields a piece proportional to the integral of $\delta(\vec{k}_t)e^{i(-k_1+k_2+k_3)\tau}$, which is fully connected but has a physical pole at $k_1=k_2+k_3$. On the other hand, consider the contraction $$\langle \zeta_{k_1} a^\dagger_{-Q}\rangle\langle\zeta_{k_2}\zeta'_{p_2}\rangle\langle\zeta_{k_3}\zeta'_{p_3}\rangle\langle a_P\zeta'_{p_1}\rangle\langle a_{-P} a^\dagger_Q\rangle$$ This piece yields again a connected result(proportional to $\delta(\vec{k}_t)$), but this time with a total energy pole at $(k_1+k_2+k_3)$.
   
   Thus, with this way of calculating things, we see we that we obtain both total and physical energy poles, but their physical meaning is obscured because of the fact that some of these might be disconnected. This diagrammatic interpretation in terms of connected/disconnected diagrams is obviously a handicap because of the unclear particle interpretation in curved spaces-in particular, we could use a different basis to expand the $\zeta$s and the resulting calculations would be a mess. One would hope that all such calculations are ultimately identical, but it is far from obvious at this stage.
   
   But it is precisely this freedom of choosing a particle interpretation that rescues us. If the initial state is chosen to be among a family of Bogolyubov transforms of the Bunch Davies states, then there is a natural choice of the mode functions where the action of the fields on the state is simple. For example, if the initial state is an $\alpha$ vacuum, it would be natural to use the mode expansion with $\alpha $ modes. This approach should be free of any artificial disconnected diagrams, because the initial state is now a vacuum state for an observer using this basis. Note that the field itself is fully invariant-observers are distinguished by their choice of mode functions and accompanying creation operators. Using the freedom of being able to choose any frame, we use the frame where \textit{the initial state is a vacuum for the choice of observer}, and then we are left with fully connected pieces. With this approach, we can expand our field using the $\alpha$ mode functions $v_k=\alpha u_k^*e^{ik\tau}+\beta u_ke^{-ik\tau}$ with corresponding modifications to the creation operators via the Bugolyubov transform, and the calculation is extremely simple because the action of these on the external states is simple.
   
   With this choice, we find
    the new mode functions, which we now call $u_k$
\begin{equation}
    u_{k}(\tau)=\alpha(1-ik\tau)e^{ik\tau}+\beta(1+ik\tau)e^{-ik\tau}
\end{equation}
where, $\alpha$, $\beta$ are complex constants. These vacua parametrized by $\alpha$ and $\beta$ are invariant under de-Sitter isometries\cite{PhysRevD.32.3136}. In other words, all co-moving/freefalling observers will agree on the choice of $\alpha$, $\beta$. 

The free two point ($O(\lambda^{0})$) function is given by
\begin{align}
    \langle\zeta_{k_1}\zeta_{k_2}\rangle=A\frac{\Delta^{2}_{\zeta}}{k_1^{3}}\delta^{3}(\vec{k_1}+\vec{k_2})
    \end{align}
    where, $A=|\alpha+\beta|^{2}$
    
    We have at $O(\lambda)$ (all non-zero contractions are equal)
    \begin{equation}
    \begin{split}
         =\textit{Re}\left(-2i\delta^{3}(\vec{k_1}+\vec{k_2}+\vec{k_3})(\alpha+\beta)^{3}\frac{\Delta^{6}_{\zeta}H^{-1}}{k_1k_2k_3}\int_{-\infty}^{0} d\tau' \tau'^{2}\left(\alpha^{*}e^{-ik_1\tau'}+\beta^{*}e^{ik_1\tau'}\right) \right.\\
         \times\left(\alpha^{*}e^{-ik_2\tau'}+\beta^{*}e^{ik_2\tau'}\right)\left(\alpha^{*}e^{-ik_3\tau'}+\beta^{*}e^{ik_3\tau'}\right)
          \end{split}
    \end{equation}
    \begin{align}
    \begin{split}
        =4\lambda\delta^{3}(\vec{k_1}+\vec{k_2}+\vec{k_3})(\alpha+\beta)^{3}\frac{\Delta^{6}_{\zeta}H^{-1}}{2k_1k_2k_3}
        \left[\left(\beta^{* \, 2}\alpha^{*}-\alpha^{* \, 2}\beta^{*}\right)\left(\frac{1}{(k_1+k_2-k_3)^{3}}+
        \frac{1}{(k_1-k_2-k_3)^{3}}\right. \right.\\
        \left.\left.+\frac{1}{(-k_1+k_2-k_3)^{3}}\right)+\left(\beta^{*3}-\alpha^{* \, 3}\right)\frac{1}{(k_1+k_2+k_3)^{3}}\right]+\textit{c.c}
         \end{split}
    \end{align}
  In the limit where $\beta=0$ i.e we are back to Bunch-Davies vacuum, the physical poles vanish therefore everything is consistent. Notice that there are by definition no $\alpha$ particles in $\alpha$ vacuum but still the three-point function exhibits physical poles. Therefore, the particle interpretation for physical poles \cite{Green:2020whw} does not hold in this case. It seems that the particle interpretation is only clear when the observer is using a  Bunch-Davies basis.

  \subsection{Coherent States}
There is yet another special class of excited states-the so called coherent states which are eigenstates for annihilation operators. Since this is an observer dependent statement in time dependent backgrounds, it is interesting to ask what poles these will generate, and will the answer change if they are defined with respect to different observers. In fact, as coherent states in quantum mechanics are known to show classical behaviour( see appendix~\ref{sec:review} for a review), their pole structure will be extremely instructive to sharpen our understanding of classical vs quantum initial states. It turns out that coherent states for a Bunch Davies observer with certain assumptions can have absence of physical poles \cite{Kundu_2012}(in contrast with \cite{Green:2020whw}). We do the same calculation with a $\frac{\lambda}{3!}\dot{\zeta^{3}}$.This is an outlier case which goes against currently accepted correspondence between initial state and pole structure. \\The coherent states are simply eigenstates of the lowering operator
 \begin{equation}
a_{\vec{k}}\ket{C}=C(\vec{k})\ket{C}
 \end{equation}
 The calculation of the Bispectrum is done again using a $\frac{\lambda}{3!}\dot{\zeta}^{3}$ with one additional assumption i.e
 \begin{equation}\label{5.2}
     C(\vec{k})+C^{*}(-\vec{k})=0
 \end{equation}
 Which is equivalent to setting the one point function to zero at $\tau=0$, that is $\langle C|\zeta(0)|C\rangle=0$. This can always be done at one point in time, and in quantum mechanics it just reflects the choice of phase of the eigenvalue of the coherent state. We need to compute at $O(\lambda)$
 \begin{equation}
     \bra{C}\zeta_{k_1}\zeta_{k_2}\zeta_{k_3}\ket{C}
 \end{equation}
 We can compute it in the following way. By wick's theorem
  \begin{equation}\label{5.4}
      \zeta  \zeta  \zeta....\zeta=:  \zeta  \zeta  \zeta....\zeta:+ :\textit{all possible contractions}:
  \end{equation}
 Since, now we are not computing in the vacuum state therefore all terms on the RHS of \eqref{5.4} contribute but the condition \eqref{5.2} makes everything vanish except the term where all fields are contracted and we get the usual BD answer. A simple proof is as follows. In every term that is not fully contracted, there exists, say, an uncontracted $a^\dagger$. There must also be a term which is different from this only by the replacement $a^\dagger\to a$. Adding these terms gives a result proportional to $C(\vec{k})+C^*(-\vec{k})$, which vanishes by assumption.  To illustrate it further, consider a term with just two contractions 
 \begin{align}\wick{
     \sim\lambda\bra{C}:\c1\zeta_{k_1}\c2\zeta_{k_2}\zeta_{k_3}\c1\zeta'_{q_1}(\tau)\c2\zeta'_{q_2}(\tau)\zeta'_{q_3}(\tau):\ket{C}}
 \end{align}
 where, we have omitted the overall integral and the overall delta function. This simplifies to
 \begin{align}\begin{split}\wick{
     \sim \lambda\c1 \zeta_{k_1}\c1\zeta'_{q_1}(\tau)\c2\zeta_{k_2}\c2\zeta'_{q_2}(\tau)}\bra{C}a_{k_3}^{\dagger}a_{q_3}^{\dagger}u_{k_3}u'_{q_3}(\tau)+a_{k_3}^{\dagger}a_{-q_3}u_{k_3}u'^{*}_{q_3}(\tau)+a_{q_3}^{\dagger}a_{-k_3}u^{*}_{k_3}u'_{q_3}(\tau)\\+a_{-k_3}a_{-q_3}u^{*}_{k_3}u'^{*}_{q_3}(\tau)\ket{C}\end{split}
 \end{align}
 where, $u_{k}$'s are the de Sitter mode functions at $\tau=0$. Using the fact that $a_{k}\ket{C}=C(k)\ket{C}$, we get
 \begin{align}
 \begin{split}\wick{
     \sim \lambda\c1 \zeta_{k_1}\c1\zeta'_{q_1}(\tau)\c2\zeta_{k_2}\c2\zeta'_{q_2}(\tau)}\bra{C}u'_{q_3}\left(C^{*}(\vec{k_3})+C(-\vec{k_3})\right)C^{*}(\vec{q_3})+u^{*}_{q_3}\left(C^{*}(\vec{k_3})\right.\\\left.+C(-\vec{k_3})\right)C(-\vec{q_3})\ket{C}\end{split}
 \end{align}
 Finally, using $C^{*}(-\vec{k})+C(\vec{k})=0$ one can easily check that the expression vanishes. In this way all partially contracted terms can be shown to vanish. Schematically, representing uncontracted operators in red, terms can be pairwise combined and cancelled as follows
\begin{equation}
 \langle C|a^\dagger a^\dagger..{\color{red}a^\dagger_K}..aaa|C\rangle+\langle C|a^\dagger a^\dagger..{\color{red}a_K}..aaa|C\rangle\propto (C(K)+C^*(-K))=0
\end{equation}
 thereby leaving behind only the fully contracted piece,
 \begin{align}
\begin{split}
    \text{Re}-2i\int\left(\prod_{i}d^{3}\vec{q}_{i}\right)\delta^{3}(\vec{k_1}
    +\vec{q_1})\delta^{3}(\vec{k_2}+\vec{q_2})\delta^{3}(\vec{k_3}+\vec{q_3})\delta^{3}(\vec{q_1}+\vec{q_2}+\vec{q_3})\\\frac{q^{2}_1q^{2}_2q^{2}_3\Delta^{6}_{\zeta}H^{-1}\lambda}{\sqrt{(k_1k_2k_3q_1q_2q_3)^{3}}}\int_{-\infty}^{0} d\tau' \tau'^{2}e^{i(q_1+q_2+q_3)\tau'}
    \end{split}
\end{align}
\begin{equation}
    =\text{Re}-2i\delta^{3}(\vec{k_1}+\vec{k_2}+\vec{k_3})\frac{\Delta^{6}_{\zeta}H^{-1}\lambda}{k_1k_2k_3}\int_{-\infty}^{0} d\tau' \tau'^{2}e^{i(k_1+k_2+k_3)\tau'}
\end{equation}
 which then yields a result identical to the Bunch Davies case upto the normalization $\langle C|\rangle$, which drops out of the overall expectation value. Thus,
 \begin{equation}
      \bra{C}\zeta_{k_1}\zeta_{k_2}\zeta_{k_3}\ket{C} =\frac{4H^{-1}\Delta^{6}_{\zeta}\lambda}{(k_1+k_2+k_3)^{3}k_1k_2k_3}\delta^{3}(\vec{k_1}+\vec{k_2}+\vec{k_3})\end{equation} \label{eq:coherent}
 This results seems to contradict the conclusion of \cite{Green:2020whw} that a total energy pole will uniquely single out BD Vacuum as the initial state. The above class of coherent states, as seen above, also shares the same feature. One can think of the physical poles vanishing precisely because of the normal ordered term vanishing. As we saw in earlier examples, physical poles occur when one two the field insertions interact with an external state at a vertex, and the third spectates. There are 2 ways to contract with the external state, either with the left or with the right. With a coherent state, each is accompanied with a factor of either $C(k)$ or $C^*(-k)$, and their sum vanishes because of our assumption. In fact, note that the vacuum state is merely a trivial special case of a coherent state with $C(\vec{k})=0$. Therefore, one can invert the train of thought and say that the vacuum state must have the same pole structure as that of a coherent state (with vanishing one point function), which implies a total energy pole for Bunch-Davies vacuum.
 
 Since the coherent state's definition is observer dependent-more precisely, dependent on the basis of annihilation/ creation operators used; one can ask what happens if the initial state is coherent not for a Bunch Davies observer but for an observer that is related by a Bogolyubov transform, say the $\alpha$-vacuum. We therefore now consider 
 
 \begin{equation}
     b_{\vec{k}}|C\rangle_\alpha=C(\vec{k})|C\rangle_\alpha
 \end{equation}
 
 This is not coherent for a Bunch Davies observer because $b\sim a-a^\dagger$ and the second piece renders it to be not an eigenstate of both $b$ and $a$. Therefore, it is a generic excited state for a Bunch Davies observer, and one would expect to see physical poles. Indeed, by expanding the field insertions with an $\alpha$ observer's mode functions, one finds

 \begin{multline}
        _\alpha\langle C|\zeta_{k_1}\zeta_{k_2}\zeta_{k_3}|C\rangle_\alpha =4\delta^{3}(\vec{k_1}+\vec{k_2}+\vec{k_3})(\alpha+\beta)^{3}\frac{\Delta^{6}_{\zeta}H^{-1}\lambda}{2k_1k_2k_3}\bigg[(\beta^{*2}\alpha^{*}-\alpha^{*2}\beta^{*})\\
         \left(\frac{1}{(k_1+k_2-k_3)^{3}}+\frac{1}{(k_1-k_2-k_3)^{3}} \right.
         \\
          \left. +\frac{1}{(-k_1+k_2-k_3)^{3}}\right)+(\beta^{*3}-\alpha^{*3})\frac{1}{(k_1+k_2+k_3)^{3}}\bigg]+\textit{c.c}
          \end{multline}
 
 The normal ordered product of the $b$'s vanishes for the same reason as before. We therefore see plainly physical poles, and this is in tune with our intuition that a state that is coherent for an $\alpha$ observer is just some generic excited state for a BD observer and therefore must display physical poles.

     \subsection{BD modes, one particle state}
\label{oneparticleBD}

 The purpose of this section is to see the precise origin of physical poles starting from a single excitation, and their support in phase space

\subsubsection{Two point function}

Performing the contractions one gets at $O(\lambda^{0})$
\begin{equation}
    =\frac{\Delta^{2}_{\zeta}}{k^{3}}\delta^{3}(\vec{k_1}+\vec{k_2})\left(1+\frac{1}{V}\left[\delta^{3}(\vec{k_1}+\vec{p})+\delta^{3}(\vec{k_1}-\vec{p})\right]\right)
\end{equation}
The $k$ dependence does not change.
\subsubsection{Three point function}
At $O(\lambda^{0})$ the correlator vanishes. Therefore, we look at $O(\lambda)$ contribution, where a typical contraction gives terms like
\begin{align}
\begin{split}
    -2i\int\left(\prod_{i}d^{3}\vec{q}_{i}\right)\delta^{3}(\vec{k_1}-\vec{p})\delta^{3}(\vec{q_3}-\vec{p})\delta^{3}(\vec{k_2}+\vec{q_1})\delta^{3}(\vec{k_3}+\vec{q_2})\delta^{3}(\vec{q_1}+\vec{q_2}+\vec{q_3})\\\frac{q^{2}_1q^{2}_2q^{2}_3\Delta^{6}_{\zeta}H^{-1}\lambda}{\sqrt{(k_1k_2k_3q_1q_2q_3)^{3}}}\int_{-\infty}^{0} d\tau' \tau'^{2}e^{-i(q_1+q_2+q_3)\tau'}
    \end{split}
\end{align}
\begin{equation}
    =-2i\delta^{3}(\vec{k_1}-\vec{p})\delta^{3}(\vec{k_1}+\vec{k_2}+\vec{k_3})\frac{\Delta^{6}_{\zeta}H^{-1}\lambda}{k_1k_2k_3}\int_{-\infty}^{0} d\tau' \tau'^{2}e^{-i(k_1+k_2+k_3)\tau'}
\end{equation}
Adding contributions from all contractions we get
%\begin{equation}
\begin{eqnarray}
     &=\frac{-4\lambda\Delta^{6}_{\zeta}H^{-1}}{k_1k_2k_3}\delta^{3}(\vec{k_1}+\vec{k_2}+\vec{k_3})\left[\left(\frac{1}{k_1+k_2+k_3}\right) \right. \nonumber \\
     & \left.+\frac{1}{V}\left(\frac{1}{(k_1+k_2+k_3)}\delta^{3}(\vec{k_1}-\vec{p})+\frac{1}{(-k_1+k_2+k_3)}\delta^{3}(\vec{k_1}+\vec{p})+k_1\leftrightarrow k_2+k_1\leftrightarrow k_3\right)\right]
    \end{eqnarray}\label{eq:disc0}
%\end{equation}
Since this is real therefore the real part is the above expression itself. Apart from a total momentum conserving delta function in (6.2) and (6.7) we also have additional delta functions involving $\vec{p}$. Therefore, this is a disconnected diagram which contributes only at isolated points in phase space, where there is an additional $V$ from the numerator that removes the delta function singularity in the denominator. The total energy pole always survives, and reflects the fact that the vacuum fluctuations can always annihilate at a vertex among themselves without any reference to initial state particles.

 Let us carefully analyse this expression in light of the discussion in literature \cite{Green:2020whw,Green:2022fwg} where the presence  of  physical poles was ascribed to physical scatterings and decays in the initial- possibly 'classical'-state. Note that there are indeed physical poles that correspond to scattering-like processes in the initial state(wherein the support of the disconnected pieces depends on the initial state momentum, as required by momentum conservation in any process), as well as a total energy pole that corresponds to vacuum fluctuations. Note that now if we divide by the norm of the state, the only unsuppressed contribution is the last term, which has the same structure as a Bunch Davies vacuum 3 point function. It is true that formally there are physical poles that when extrapolated to flat space would indicate interactions between the initial state particle and the field insertions-what this means is that there exist diagrams with interaction vertices which can connect the external $p$ state with two of the $k_i$ modes. The remaining $p$ state is then contracted with the remaining field insertion, and momentum conservation produces the offending delta function that makes the support localized. Dividing by the norm then forces this term to always vanish, except when the support is reached i.e. the unique point in phase space where one of the field insertions has the same momentum as the initial state. It is natural to ask what happens when the initial state contains all possible momenta instead of a single particle with a fixed momentum $p$. The physical pole should now be more likely to be observable, because of the larger support. In fact, it was shown in \cite{Martin:1999fa} that an initial state with a finite number of quanta is already at odds with data(which is largely compatible with Bunch Davies initial conditions), and therefore it is natural to ask how this changes when there is a very large number of quanta in the initial state-atleast one for every momentum. In this case, the isolated support for the physical poles is expected to smear out, because there always be a momentum in the initial state that has support at the same momentum as one of the $k_i$. Then, while the physical poles will still formally have support at isolated momenta, the fact that the phase space of support is continuous will ensure that atleast one of the physical poles is always reached. Let us see how this happens explicitly. Consider, an initial state where multiple momenta are occupied; $\sqrt{2p_1\cdot 2p_2\cdot 2p_3\cdot\cdot\cdot}|p_1,p_2,p_3,....\rangle$. This state will be normalized with respect to $\prod\langle p_i|p_i\rangle$; all other contractions are zero by orthogonality. Then, the 3 point function will be schematically-

\begin{multline}
  \frac{\langle p_1,p_2,p_3,...|\zeta_{k_1}\zeta_{k_2}\zeta_{k_3}|p_1,p_2,p_3,..\rangle}{\prod\langle p_i|p_i\rangle}\propto\delta(\vec{k_t})\bigg[\frac{1}{(k_1+k_2+k_3)^3}\\
  \\+\frac{1}{(k_1+k_2+k_3)^3}\sum_i\frac{\delta(\vec{k}_1-\vec{p}_i)}{\langle p_i|p_i\rangle}+\frac{1}{(k_2+k_3-k_1)^3}\sum_i\frac{\delta(\vec{k}_1+\vec{p}_i)}{\langle p_i|p_i\rangle}+\textrm{perm.}\bigg]
\end{multline}\label{eq:disc1}

All other contractions vanish because they involve spectator terms like $\delta(\vec{p}_i-\vec{p}_j)$ which for unequal momenta vanish. Note now that in comparison to the 1 particle state, the physical poles have more support - at each of the $p_i$s now, and the observational imprint of these states. If the momenta were continuously occupied-that is to say that $p_i$s were a continuous variable, then we would expect a continuous distribution. Therefore, taking the continuum limit and making the replacement

\begin{equation}
    \sum_i \frac{ \delta(\vec{k}_1\pm\vec{p}_i)}{\langle p_i|p_i\rangle}\to\sum_{\vec{p}\in \mathbb{R}^3}\frac{ \delta(\vec{k}_1\pm\vec{p})}{\langle p|p\rangle}=1
\end{equation}\label{eq:disc2}

because the sum now has all possible momenta, one of which must be $k_1$ and therefore this term always has a support. The numerator and denominator are then both $\delta^{(3)}(0)$ which can be regulated in the usual way \cite{Schwartz:2014sze} to be taken to $V$, and therefore the limit $V\to\infty$ of this expression is just $1$.This is similar to how the infinities coming from phase space volumes drop out in the final formulae for cross sections.
One can think of this infinite sum schematically as an integral $\int d^3k \delta(\vec{k}_1\pm\vec{p})$=1, where it is understood that in the denominator the limit $d^3p\to 0, V\to \infty$ is taken such that the $d^3p V\to \textrm{finite}$.

We therefore see that in the case where \textit{all} momenta states are occupied, the physical pole is always reached. While it is true that it comes from a disconnected diagram, the point now is that \textit{now there always exists a disconnected diagram that has the required support to be observable}. We can therefore conclude that

\begin{equation}
    \textrm{initial state}=\prod_{k\in \mathbb{R}^3}|\vec{k}\rangle\implies \textrm{sum of disconnected pieces always contains observable physical pole}
\end{equation}

It turns out that if one wants a connected contribution from $\ket{p}$ state correlation function then one needs to go to higher order in perturbation theory. e.g., for our case, we need to compute contribution at $O(\lambda^{3})$. The in-in formula at $O(\lambda^{3})$ gives
\begin{equation}
 -i\int_{-\infty}^{0}dt_{1}\int_{-\infty}^{t_1}dt_2\int_{-\infty}^{t_2}dt_3[H(t_3),[H(t_2),[H(t_1),Q(0)]]]
\end{equation}
 Clearly, there will be many terms but it serves our purpose to just pick one of these terms and one of the contractions. We pick
 \begin{align} \begin{split}
     -2\textit{Re}\int_{-\infty}^{0}\frac{d\tau_{1}}{H\tau_1}\int_{-\infty}^{\tau_1}\frac{d\tau_2}{H\tau_2}\int_{-\infty}^{\tau_2}\frac{d\tau_3}{H\tau_{3}}i\langle Q(0)H(t_1)H(t_2)H(t_3)\rangle\\
    =-2\textit{Re}\int_{-\infty}^{0}dt_{1}\int_{-\infty}^{t_1}dt_2\int_{-\infty}^{t_2}dt_3i\prod_{i=1}^{9}\int d^{3}\Vec{q_{i}}\delta^{3}(\vec{q_1}+\vec{q_2}+\vec{q_3})\delta^{3}(\vec{q_4}+\vec{q_5}+\vec{q_6})\\\times\delta^{3}(\vec{q_7}+\vec{q_8}+\vec{q_9})\bra{p}\zeta_{k_1}\zeta_{k_2}\zeta_{k_3}\zeta'_{q_1}(\tau_1)\zeta'_{q_2}(\tau_1)\zeta'_{q_3}(\tau_1)\zeta'_{q_4}(\tau_2)\\\times\zeta'_{q_5}(\tau_2) \zeta'_{q_6}(\tau_2)\zeta'_{q_7}(\tau_3)\zeta'_{q_8}(\tau_3)\zeta'_{q_9}(\tau_3)\ket{p}
      \end{split}
 \end{align}
 It can be easily checked that if one performs the following contraction
 \begin{equation}
     \langle a_{p}\zeta_{q_1}(\tau_1)\rangle     \langle\zeta_{q_2}(\tau_1)\zeta_{q_4}(\tau_2)\rangle     \langle\zeta_{q_5}(\tau_2)\zeta_{q_7}(\tau_3)\rangle     \langle\zeta_{k_1}\zeta_{q_3}(\tau_1)\rangle     \langle\zeta_{k_2}\zeta_{q_6}(\tau_2)\rangle     \langle\zeta_{k_3}\zeta_{q_9}(\tau_3)\rangle     \langle\zeta_{q_8}(\tau_3)a_{p}^{\dagger}\rangle
 \end{equation}
  then after performing all momenta integrals one is left with only one delta function i.e $\delta^{3}(\vec{k_1}+\vec{k_2}+\vec{k_3})$ which means we have a connected diagram. In the next section, we will see in detail the nature of contractions for a highly occupied initial state, by which we mean not only that all momenta are excited, but also each is excited by a large number. In Appendix \ref{sec:alpha1particle}, we show that if instead we have a 1-particle state for an $\alpha$ observer, we see physical poles without isolated support.
 
\subsection{BD modes, large occupation number initial state}
Following\cite{Green:2022fwg} and the preceding discussion, we compute the correlation functions for excited high occupation number states above the Bunch-Davies vacuum. All momentum modes for this $\ket{n}$ state are highly occupied. To be more precise we have
\begin{equation}
    \ket{n}\equiv\bigotimes_{\vec{k_{i}}}\ket{n_{\vec{k}}}
\end{equation}
where,
\begin{equation}
    \ket{n_{\vec{k}}}=\frac{1}{\sqrt{n!}}\left(a_{\vec{k}^{\dagger}}\right)^{n}\ket{0}
\end{equation}
where $n$ is very large.
\subsubsection{Two-Point Function}
The two point function at $O(\lambda^{0})$ is given by
\begin{align}
    \bra{n}\zeta_{k_1}\zeta_{k_2}\ket{n}=(2\pi)^{3}\delta^{3}(\vec{k_1}+\vec{k_2})\left((n_{k_1}+1)u^{*}_{k_2}u_{k_1}+n_{k_1}u_{k_2}u^{*}_{k_1}\right)\\
    =(2pi)^{3}\delta^{3}(\vec{k_1}+\vec{k_2})\frac{\Delta^{2}_{\zeta}}{k_1^{3}}(2n_{k}+1)
\end{align}
where, the action of field $\zeta_k$ on the state is given as follows
\begin{equation}
    \zeta_{k}\ket{n}=\sqrt{n_{-k}+1}u_{k_1}\ket{n_{-k_1}+1}\ket{\hat{n};-\vec{k_{1}}}+\sqrt{n_{k}}u^{*}_{k}\ket{n_{k}-1}\ket{\hat{n};\vec{k}}
\end{equation}
\subsubsection{Three-Point Function}
\label{4.3.2}
At $O(\lambda)$ the three point function is the following
\begin{align}
      \textit{Re}\bra{0}\prod_{i}(a_{p_i})^{n}\left(-2i\zeta^{I}_{k_1}\zeta^{I}_{k_2}\zeta^{I}_{k_3}\int\left(\prod_{i}d^{3}\vec{q}_{i}\right)\delta^{3}(\sum \vec{q_{i}})\int_{-\infty}^{0}\frac{d\tau'}{H\tau'}\left(\zeta'^{I}_{q_1}\zeta'^{I}_{q_2}\zeta'^{I}_{q_3}\right)\right)\prod_{i}(a^{\dagger}_{p_i})^{n}\ket{0}
\end{align}
It is clear from the above expression and previous computations that there are contraction where $\zeta'_{q}$s on the right can be contracted with lowering operators on the left and the $\zeta_{k}$s on the left with raising operators on the right. These types of contractions will give rise to physical poles. One such contraction is shown below
\begin{equation}
\wick{
    \prod_{i}\c1(a_{p_i})^{n}\c2a_{p_{j}}(-2i  \c3\zeta^{I}_{k_1}\c4\zeta^{I}_{k_2}\c5\zeta^{I}_{k_3}\int\left(\prod_{i}d^{3}\vec{q}_{i}\right)\delta^{3}(\sum \vec{q_{i}})\int_{-\infty}^{0}\frac{d\tau'}{H\tau'}\left(\c5\zeta'^{I}_{q_1}\c4\zeta'^{I}_{q_2}\c2\zeta'^{I}_{q_3})\right)\c3 a^{\dagger}_{p_{j}}\prod_{i}\c1(a^{\dagger}_{p_i})^{n}
    }
\end{equation}
Such a contraction will produce an integral of the form
\begin{align}
    \sim \int_{-\infty}^{0}\tau'^{2}e^{-i\left(k_2+k_3-k_1\right)\tau}d\tau'
\end{align}
giving rise to physical poles but again there will be disconnected diagrams. 

\vspace{5mm}

 \subsection{Four point function} \label{sec:fourpt}
      All these calculations assume that there is no nongaussianity in the initial state. Then the Nongaussianity results only from the interaction Hamiltonian time evolving  the initial state to the present state. Moreover, one can explicitly write the BD interacting vacuum in terms of the free theory eigenstate using standard perturbation theory, and Trivedi shows that the limit $\tau_{initial}\to -\infty$ reduces the interacting theory vacuum to free vacuum only for cubic self coupling. Our analysis has more possibilities, where the initial state needn't match the vacuum of the mode functions, and therefore really has no access to convenient methods like the WKB approximation to interpret results. We aim to only study the analytic structures allowed, and in this section we extend the results of \cite{Green:2020whw} to four-point functions\cite{Goodhew_2021,Arroja_2009,Chen_2009,https://doi.org/10.48550/arxiv.2106.05294}. We calculate both classical and quantum four-point functions and comment about their pole structures.\\
\subsubsection{Classical correlator}
 
 Consider a massless field in de Sitter space with the following action
\begin{equation}
    S=\int{d\tau d^{3}x\left(\frac{1}{2}\left(a^{2}\zeta'^2-a^2\partial^2\zeta\right)+a\frac{\lambda}{6}\zeta'^3\right)}
\end{equation}
where $a$ is the scale factor and prime denotes a derivative w r t to conformal time $\tau$.
The equation of motion is given by
\begin{equation}
    \frac{1}{a^2}(\zeta''+2\frac{a'}{a}\zeta'-\partial^2\zeta)=-\frac{\lambda}{a^4}(a\zeta'\zeta''+\frac{a'\zeta'^2}{2})
\end{equation}
The Greens function corresponding to the operator on the left acting on $\zeta$ and its derivative are given by\cite{Green:2020whw,van_der_Meulen_2007}

\begin{align}
     G_{\vec{k}}(\tau,\tau')=\theta(\tau-\tau')\frac{2\Delta_{\zeta}^2}{k^3}\left((1+k^2\tau\tau')\sin{k(\tau-\tau')}-k(\tau-\tau')\cos{k(\tau-\tau')}\right)\\
     \partial_{\tau}G_{\vec{k}}(\tau,\tau')=2\Delta_{\zeta}^2[k^{-1}\tau'\sin{k(\tau-\tau')}-\tau\tau'\cos{k(\tau-\tau')}]
\end{align}
We can solve Eq(5.2) perturbatively by expanding $\zeta$ as follows
\begin{equation}
    \zeta=\zeta^{(1)}+\lambda\zeta^{(2)}+\lambda^2\zeta^{(3)}+O(\lambda^3)
\end{equation}
Plugging this into (5.2) we get the following equations 

\begin{align}
    \frac{1}{a^2}({\zeta''}^{(1)}+2\frac{a'}{a}{\zeta'}^{(1)}-\partial^2\zeta^{(1)})=0 \\
      \frac{1}{a^2}({\zeta''}^{(2)}+2\frac{a'}{a}{\zeta'}^{(2)}-\partial^2\zeta^{(2)})+\frac{1}{a^4}\left(a{\zeta'}^{(1)}{\zeta''}^{(1)}+\frac{a'{\zeta'}^{(1)2}}{2}\right)=0\\
      \frac{1}{a^2}({\zeta''}^{(3)}+2\frac{a'}{a}{\zeta'}^{(3)}-\partial^2\zeta^{(3)})+\frac{1}{a^4}\left(a{\zeta'}^{(1)}{\zeta''}^{(2)}+a{\zeta'}^{(2)}{\zeta''}^{(1)}+a'{\zeta'}^{(1)}{\zeta'}^{(2)}\right)=0
\end{align}
Therefore the solution to (5.7) and (5.8) can be written in terms of Greens function given above. 
\begin{align}
    \zeta^{(2)}(\tau,\vec{x})=\int d^{3}x'\int{d\tau'\frac{ a}{2}\partial_{\tau'}G(x,x')(\zeta'^{(1)2})}\\
    \zeta^{(3)}(\tau,\vec{x})=\int d^{3}x\int{d\tau'a\left(\partial_{\tau'}G(x,x')\zeta'^{(1)}\zeta'^{(2)}\right)}
\end{align}
where we have neglected a total derivative since $G(\tau,\tau+0^{+})$ is zero (retarded GF) and $G(\tau,-\infty)$ is also set to zero with an appropriate $i\epsilon$ prescription.
These equations in fourier space read 
\begin{align}
 \zeta^{(2)}_{\vec{k}}(\tau)=\int{d\tau'\frac{d^{3}p}{2(2\pi)^{3}}a\left(\partial_{\tau'}G_{\vec{k}}(\tau,\tau'){\zeta'}^{(1)}_{\vec{p}}{\zeta'}^{(1)}_{\vec{k}-\vec{p}}\right)}\\
     \zeta^{(3)}_{\vec{k}}(\tau)=\int{d\tau'\frac{d^{3}p}{(2\pi)^{3}}a\left(\partial_{\tau'}G_{\vec{k}}(\tau,\tau'){\zeta'}^{(2)}_{\vec{p}}{\zeta'}^{(1)}_{\vec{k}-\vec{p}}\right)}
\end{align}
The free field solution is given by 
\begin{equation}
    \zeta^{(1)}_{\vec{k}(\tau)}=\frac{\Delta_{\zeta}}{\sqrt{k^{3}}}\left( a_{\vec{k}}^{\dagger}u_{k}(\tau)+a_{-\vec{k}}u_{k}^{*}(\tau)\right)
\end{equation}
defined with the following statistics 
\begin{align}
    <a_{\vec{k}}a_{\vec{k'}}^{\dagger}>= <a_{\vec{k}}^{\dagger}a_{\vec{k'}}>=\frac{1}{2}\delta^{3}(\vec{k}-\vec{k'})
\end{align}
where, $u_{\vec{k}}(\tau)=(1-ik\tau)e^{ik\tau}$\\
This produces the correct two point function which is the only observable we have measured so far. Note that this statistical nature is due to our lack of knowledge and not due to some inherent uncertainty as is the case for quantum mechanical operators.\par
The two point function is given by 
\begin{equation}
    <\zeta^{(1)}_{\vec{k}}(\tau)\zeta^{(1)}_{\vec{k'}}(\tau')>=\frac{\Delta_{\zeta}^{2}}{k^{3}}\delta^{3}(\vec{k}+\vec{k'})Re[u_{k}(\tau)u_{k}^{*}(\tau')]
\end{equation}\\
Let us now calculate the classical 4 point function  upto $O(\lambda^{2})$
\begin{align}
\begin{split}
    <\zeta\zeta\zeta\zeta>=<(\zeta^{(1)}_{k_1}+\lambda\zeta^{(2)}_{k_1}+\lambda^{2}\zeta^{(3)}_{k_1})(\zeta^{(1)}_{k_2}+\lambda\zeta^{(2)}_{k_2}+\lambda^{2}\zeta^{(3)}_{k_2})(\zeta^{(1)}_{k_3}+\lambda\zeta^{(2)}_{k_3}+\lambda^{2}\zeta^{(3)}_{k_3})\times\\(\zeta^{(1)}_{k_4}+\lambda\zeta^{(2)}_{k_4}+\lambda^{2}\zeta^{(3)}_{k_4})>
    \end{split}
\end{align}
Terms of $O(\lambda)$ vanish (odd number of terms) and the non-trivial contribution to the correlator comes from two types of terms:\\\\
\textbf{I}: $<\zeta^{(1)}_{k_1}\zeta^{(1)}_{k_2}\zeta^{(2)}_{k_3}\zeta^{(2)}_{k_4}>$\\\\
\textbf{II}: $<\zeta^{(1)}_{k_1}\zeta^{(1)}_{k_2}\zeta^{(1)}_{k_3}\zeta^{(3)}_{k_4}>$\\\\
Rest of the terms are just permutations of these two terms.\par
Let us first evaluate \textbf{I}:

\begin{align}
\begin{split}
    \textbf{I}=\int\frac{d\tau'd\tau''d^{3}pd^{3}q}{(2\pi)^{6}}(-H\tau')^{-1}(-H\tau'')^{-1}\partial_{\tau'}G_{k_3}(\tau,\tau')\partial_{\tau'}G_{k_3}(\tau,\tau'')\times\\<\zeta_{k_1}^{(1)}\zeta_{k_2}^{(1)}{\zeta'}_{\vec{k_3}-\vec{p}}^{(1)}(\tau'){\zeta'}_{p}^{(1)}(\tau'){\zeta'}_{\vec{k_4}-\vec{q}}^{(1)}(\tau''){\zeta'}_{q}^{(1)}(\tau'')>
    \end{split}
    \end{align}
    Using wick's theorem for these gaussian random fields, we get
    \begin{align}
    \begin{split}
     =\frac{12\Delta^{10}_{\zeta}H^{-2}k_{13}}{6k_1k_2k_3k_4}\left[\int d\tau' \tau'^{2}\cos{k_1\tau'}\cos{k_{13}\tau'}\sin{k_3\tau'}\int d\tau' \tau'^{2}\cos{k_2\tau''}\cos{k_{13}\tau''}\sin{k_4\tau''}\right]\\+k_1\leftrightarrow k_2
     \end{split}
\end{align}
where, $k_{13}=|\vec{k_1}+\vec{k_3}|$. After some more algebra this simplifies to
\begin{equation}\label{4.66}
    \frac{4\Delta^{10}_{\zeta}H^{-2}}{k_1k_2k_3k_4}\left(k_{13}\times\alpha\times\beta+k_1\leftrightarrow k_2\right)
\end{equation}
where, 
\begin{align}
    \alpha=\frac{1}{(k_1+k_3+k_{13})^{3}}+\frac{1}{(k_1+k_3-k_{13})^{3}}+\frac{1}{(-k_1+k_3+k_{13})^{3}}-\frac{1}{(k_1-k_3+k_{13})^{3}}\\ \beta=\frac{1}{(k_2+k_4+k_{13})^{3}}+\frac{1}{(k_2+k_4-k_{13})^{3}}+\frac{1}{(-k_2+k_4+k_{13})^{3}}-\frac{1}{(k_2-k_4+k_{13})^{3}}.
\end{align}\\\\
Now, term \textbf{II} gives
\begin{align}
\begin{split}
     = \int\frac{d\tau'd^3p}{(2\pi)^{3}}(-H\tau')^{-1}\partial_{\tau'}G_{k_4}(\tau,\tau')<\zeta_{k_1}^{(1)}\zeta_{k_2}^{(1)}\zeta_{k_3}^{(1)}{\zeta'}_{\vec{k_4}-\vec{p}}^{(1)}(\tau')\times\\\partial_{\tau'}\left( \int\frac{d\tau''d^{3}q}{(2\pi)^{3}}(-H\tau'')^{-1}\partial_{\tau''}G_{p}(\tau',\tau''){\zeta'}_{q}^{(1)}(\tau''){\zeta'}_{\vec{p}-\vec{q}}(\tau'')\right)>
     \end{split}
\end{align}

Using Leibniz rule to simply this further 
\begin{align}
\begin{split}
    \int\frac{d\tau'd^3pd^3qd\tau''}{(2\pi)^{6}}(-H\tau')^{-1}(-H\tau'')^{-1}\partial_{\tau'}G_{k_4}(\tau,\tau')\partial_{\tau'}\partial_{\tau''}G_{p}(\tau',\tau'')\\<\zeta_{k_1}^{(1)}\zeta_{k_2}^{(1)}\zeta_{k_3}^{(1)}{\zeta'}_{\vec{k_4}-\vec{p}}^{(1)}(\tau'){\zeta'}_{q}^{(1)}(\tau''){\zeta'}^{(1)}_{\vec{p}-\vec{q}}(\tau'')>
    \end{split}
\end{align}
We proceed as before and after some lengthy algebra one gets

\begin{multline}\label{4.71}
\frac{4H^{-2}\Delta^{10}_{\zeta}k_{14}}{16k_1k_2k_3k_4}\bigg[\frac{4}{u_{x1}^{3}}\bigg(\frac{1}{x^{3}}+\frac{6}{xu^{2}_{x1}}\bigg)+\frac{4}{u_{x2}^{3}}\bigg(\frac{1}{x^{3}}+\frac{6}{xu^{2}_{x2}}\bigg)+\frac{4}{u_{x3}^{3}}\bigg(\frac{1}{{x}^{3}}+\frac{6}{xu^{2}_{x3}}\bigg)\\+\frac{4}{u_{x4}^{3}}\bigg(\frac{1}{{x}^{3}}+\frac{6}{xu^{2}_{x4}}\bigg)+\frac{12}{x^{2}}\bigg(-\frac{1}{u^{4}_{x1}}+\frac{1}{u^{4}_{x2}}-\frac{1}{u^{4}_{x3}}+\frac{1}{u^{4}_{x4}}\bigg)+(u_y,y)+(u_z,z)+(u_w,w)
    \\+k_1\leftrightarrow k_2(\textrm{of previous term})+k_2\leftrightarrow k_3 (\textrm{of previous term})\bigg]
\end{multline}

where,
\begin{align}
(u_m,m)=\frac{4}{u_{m1}^{3}}\left(\frac{1}{m^{3}}+\frac{6}{mu^{2}_{m1}}\right)+\frac{4}{u_{m2}^{3}}\left(\frac{1}{m^{3}}+\frac{6}{mu^{2}_{m2}}\right)+\frac{4}{u_{m3}^{3}}\left(\frac{1}{{m}^{3}}+\frac{6}{mu^{2}_{m3}}\right)\\+\frac{4}{u_{m4}^{3}}\left(\frac{1}{{m}^{3}}+\frac{6}{mu^{2}_{m4}}\right)+\frac{12}{m^{2}}\left(-\frac{1}{u^{4}_{m1}}+\frac{1}{u^{4}_{m2}}-\frac{1}{u^{4}_{m3}}+\frac{1}{u^{4}_{m4}}\right)\\
    x=k_{14}+k_2+k_3\\y=k_{14}+k_2-k_3\\z=k_{14}-k_2+k_3\\w=k_{14}-k_2-k_3\\u_{x1}=k_1+k_4+k_{14}-x\\u_{x2}=k_1+k_4-k_{14}-x\\u_{x3}=-k_1+k_4-k_{14}-x\\u_{x4}=-k_1+k_4-k_{14}-x
\end{align}
For $u_y$ and $u_z$ terms we just replace $x$ in $u_x$ terms by $y$ and $z$. Again, the poles are located at physical momenta consistent with the interpretation that if initial state has physical particles then processes like decays can give rise to physical poles. The final expression for the classical Four-point function is the sum of \eqref{4.66} and \eqref{4.71} + permutations.

\subsubsection{Quantum correlator}

We recall the mode expansion
\begin{equation}
    \zeta_{\vec{k}}(\tau)=\frac{\Delta_{\zeta}}{\sqrt{2k^3}}(u_k(\tau)a_{\vec{k}}+u^*_k(\tau)a^\dagger_{-\vec{k}}), u_k(\tau)=(1+ik\tau)e^{-ik\tau}
\end{equation} 

From which we find

\begin{equation}
    \zeta'_k(\tau)=\frac{\Delta_{\zeta}k^2\tau}{\sqrt{2k^3}}(a_{\vec{k}}e^{-ik\tau}+a^\dagger_{-\vec{k}}e^{ik\tau})
\end{equation}

We will need the following 2 point functions-

\begin{equation}
  \langle \zeta'_{p}(\tau_1)\zeta'_{q}(\tau_2)\rangle=(2\pi)^3\delta(\vec{p}+\vec{q})\frac{p\Delta_{\zeta}^2\tau_1\tau_2}{2}e^{ip(\tau_2-\tau_1)}\equiv (2\pi)^3\delta(\vec{p}+\vec{q}) G'_p(\tau_1,\tau_2)
\end{equation}

\begin{equation}
   (2\pi)^3\delta(\vec{p}+\vec{k}) G_k(\tau)\equiv\langle \zeta'_p(\tau)\zeta_k(0)\rangle=(2\pi)^3\delta(\vec{p}+\vec{k})\frac{\Delta_{\zeta}^{2}\tau}{2k}e^{-ik\tau}=(2\pi)^3\delta(\vec{p}+\vec{k})\langle \zeta_k(0)\zeta'_p(\tau)\rangle ^*
\end{equation}

At order $\lambda^2$, we have-

\begin{align}
    \langle \zeta_{k_1}\zeta_{k_2}\zeta_{k_3}\zeta_{k_4}\rangle=-2Re\int^0_{-\infty}d\tau_2\int^{\tau_2}_{-\infty}d\tau_1 \bigg(\langle \zeta_{k_1}\zeta_{k_2}\zeta_{k_3}\zeta_{k_4}H(\tau_2)H(\tau_1)\rangle\\
    \nonumber \\-\langle H(\tau_1)\zeta_{k_1}\zeta_{k_2}\zeta_{k_3}\zeta_{k_4}H(\tau_2)\rangle\bigg)
\end{align}

We define 

\begin{equation}
    \vec{k}_{12}=\vec{k}_1+\vec{k}_2=-\vec{k}_{34}\hspace{5mm}\vec{k}_t=\sum_{i=1}^4\vec{k}_i
\end{equation}

This becomes, suppressing momentum conserving delta functions and permutations for brevity,

\begin{multline}
    = -2Re\int^0_{-\infty}\frac{d\tau_2}{H\tau_2}\int^{\tau_2}_{-\infty}\frac{d\tau_1}{H\tau_1}\bigg(G^*_{k_1}(\tau_2)G^*_{k_2}(\tau_2)G^*_{k_3}(\tau_1)G^*_{k_4}(\tau_1)G'_{k_{12}}(\tau_2,\tau_1)\\
    \\-G_{k_1}(\tau_1)G_{k_2}(\tau_1)G^*_{k_3}(\tau_2)G^*_{k_4}(\tau_2)G'_{k_{12}}(\tau_1,\tau_2)\bigg)+\textrm{perm.}
\end{multline}

\begin{multline}
    =-2Re\frac{\Delta_{\zeta}^{10}k_{12}}{32k_1k_2k_3k_4}\int^0_{-\infty}\frac{d\tau_2}{H\tau_2}\int^{\tau_2}_{-\infty}\frac{d\tau_1}{H\tau_1}\tau_1^3\tau_2^3\bigg(e^{i(k_1+k_2)\tau_2+(k_3+k_4)\tau_1+{k_{12}}(\tau_1-\tau_2)}\\
    \\-e^{-i(k_1+k_2)\tau_1+(k_3+k_4)\tau_2+{k_{12}}(\tau_2-\tau_1)}\bigg)
\end{multline}

And therefore the full 4 point function becomes, after including all possible contractions

\begin{equation}\begin{split}\frac{4\lambda^{2}\Delta^{10}_{\zeta}H^{-2}k_{12}}{k_1k_2k_3k_4}\left(\frac{2}{k_t^{3}(k_3+k_4+k_{12})^{3}}+\frac{6}{k_{t}^{4}(k_3+k_4+k_{12})^{2}}+\frac{12}{k^{5}_{t}(k_3+k_4+k_{12})}\right.\\
\\\left.+\frac{1}{(k_1+k_2+k_{12})^{3}(k_3+k_4+k_{12})^{3}}\right)+\textit{permutations}\end{split}\end{equation}

We see the presence of additional poles besides the one for total energy, but this is no physical pole, consistent with our expectations from a Bunch Davies vacuum.

 \section{The origin of singularities}
We must clarify what we aim to understand from the above calculations. We wish to probe the interplay between the nature of the initial state and the structure of in-in correlators, through the momentum dependence of the singularity structures. What we mean by this is identifying the \textit{configuration of momenta for which these correlation functions become singular}. The functional form of the singularity isn't fixed, and as we will see in a later section, depends crucially on the nature of interactions. We have assumed the initial conditions are set at $\tau=-\infty$, and in section \ref{sec:cutoff} we see how things change at a finite early time cutoff.

The authors of \cite{Green:2022fwg} studied this in the context of understanding whether the initial conditions of inflation were classical perturbations in an ensemble or quantum zero point fluctuations of the BD vacuum. The pole structure was linked to physical processes-such as particle decay in the initial (classical) state producing poles at physical momenta. However, the results of \cite{Shukla:2016bnu} exhibit the physical poles despite considering $\alpha$-mode insertions in an initial $\alpha$-vacuum. This displays how the notion of calling a state 'classical' based on the particle content is subtle, as the $\alpha$ vacuum-while excited for a Bunch Davies observer, is devoid of $\alpha$-particles. We will also see how the signatures of initial state effects  reflect in the modification to Maldacena's consistency criteria(which is already modified in the presence of higher derivative operators \cite{Creminelli:2003iq})-in particular, non-Bunch Davies initial states produce severely enhanced deviations from the consistency criteria. 

However, one needs to be careful about the physical implications of these singularity structures. We have seen the inevitability of disconnected diagrams in correlation functions. Consider a flat space correlator with 3 field insertions and at cubic order. If the external state is a single particle state of fixed momentum, then its natural contribution to the S-matrix element is schematically

\begin{equation}
    \langle p|T\phi(k_1)\phi(k_2)\phi(k_3)|p\rangle\to^{LSZ}\langle p,k_1,k_2|p,k_3\rangle
\end{equation}

where it is understood that crossed elements can be generated too; we have just written down a possible configuration. Now, at first order, it is easy to see that this amplitude will contain disconnected subgraphs where some of the particles  are mere spectators- only $3$ out of $5$ particles can interact at a cubic vertex, and the remaining two will act as spectators which don't participate in the interaction. These diagrams will produce factors of $\delta(\textrm{subset of } p,k_i)$, and will therefore only have support at isolated points in phase space. Generating fully connected contributions will require us to go to higher order in perturbation theory, where there are enough vertices to contract with every external leg-but these are subleading effects. It then becomes very tricky to find measurable imprints of these pole structures. In fact, the whole discussion is subtle because it is unclear how to translate from flat space intuition of 'particles' and their Feynman diagrams because the notion of particles is itself subtle in curved spaces. For example, it is entirely possible to switch to a different picture of particles, and the resulting correlators will have to then be interpreted differently. As an example which will be discussed in detail later, if we consider an initial state which is an excited Bunch Davies state such that it happens to be the vacuum for an $\alpha$-observer, then the correlator can be computed in two ways. We can write the state in terms of Bunch Davies excitations and compute the correlator that will now display disconnected pieces because of the initial state particles, see \ref{sec:alpha}. Alternatively, we can go to a frame where there are no excitations-this frame corresponds to choosing the $\alpha$-vacuum basis for mode expansion of our fields. Then, the diagrammatic interpretation of the involved terms will be entirely different, and we will see no disconnected pieces simply because now the interactions at a vertex don't involve initial state particles. Schematically, the two ways to compute the correlator amount to

\begin{equation}
    _\alpha\langle 0|\zeta^{\alpha}\zeta^{\alpha}\zeta^{\alpha}|0\rangle_\alpha \hspace{3mm} \textrm{or}\hspace{3mm} _{BD}\langle \big(0|e^{\int d^3p a_pa_{-p}})|\zeta^{BD}\zeta^{BD}\zeta^{BD}|\big(e^{\int d^3p a^\dagger_p a^\dagger_{-p}}|0\big)\rangle_{BD}
\end{equation}

Where we have indicated what mode functions i.e. what particle interpretation is being used. $a_p$ are Bunch Davies operators. Note that the second way of calculating requires to include contractions representing interactions of in-state particles at a vertex, whereas the first doesn't. One expects that eventually some kind of resummation renders the two approaches equivalent, but it is not obvious. 

The take-home message is this-once we begin considering excited initial states, then \textit{because only finitely many of them can interact at a finite order in perturbation theory} (determined by the nature of the coupling and the order at which we are interested), correlation functions will be plagued by disconnected diagrams. Many pole structures will be obscured in observation because of the fact that their kinematic support lies on isolated points in phase space. 

The disconnectedness of these diagrams clouds the interesting pole structure that comes from genuine interaction. If, then, we were in a situation where there are no real particles in the  initial state, then one would expect to be free from these disconnected diagrams. This is in fact what happens whenever the initial state is a vacuum state of the perturbation modes-and when this vacuum state is 'excited' relative to a Bunch Davies vacuum in a manner that all momenta are excited, one sees the presence of genuine, physically measurable physical poles.  We would like to understand, then, to what extent do pole structures encode information about the  initial state. Let us first understand their origin.

\subsection{Singularities and interacting vertices: flat space vs curved space}\label{sec:LSZ}

Let us elaborate on the remarks of the previous section, drawing analogy from flat space results to see explicitly how the pole structure for in-in correlators is different from in-out and its interpretation in terms of physical processes. We will also see the precise origin of pole structures and the role of mode mixing. Our object of interest will be the structure of kinematics at a vertex in perturbation theory, since it is integrals over these vertices that produce analytic structures.

In flat space, we typically calculate time ordered correlators via Wick's theorem. Let us first see how kinematics are constrained at a vertex in a simple $\lambda\phi^3/3!$ interaction in flat space. If we calculate the time ordered correlator, then the leading contribution to the 3 point correlator in perturbation theory is of the type

\begin{equation}
    \langle T \phi(x_1)\phi(x_2)\phi(x_3)\rangle_{int}\sim\lambda \int d^4x D_F(x_1-x)D_F(x_2-x)D_F(x_3-x)
\end{equation}

where $D_F(x)=\int \frac{d^4p}{(2\pi)^4}e^{ipx}/(p^2+i\epsilon)$ is the Feynman propagator, the use of which is tied to the presence of time ordering on the left. Let us focus on the integral over the vertex, where the momenta in the propagator's expression combine into a delta function, which tells us that four-momentum is conserved at every vertex. Since the $p_i$ are being integrated over, and noting that $D_F(x)=D_F(-x)$, it is clear that it really doesn't matter if we choose to call a momentum $p$ or $-p$. In fact, being a vertex or a correlation function, it doesn't immediately correspond to any physical process. Therefore, \textit{we need to specify additional information to specify the correct kinematic constraints for any physical process}-this is precisely the LSZ prescription. This expression depends on $x_i$-the points of insertion. While the LHS is agnostic to their relative order because of time ordering within which everything commutes; the RHS needs greater care. Let us see how LSZ systematically teaches us what to do. Let us say that the field insertions at $x_i$ correspond to one particle states of momentum $k_i$ in the asymptotic region, where the theory is assumed to be free(this is one of the assumptions that break down in curved space). For concreteness, let us say that $x_1,x_2$ correspond to insertions in the far past(in-state), and $x_3$ to an insertion in the far future(out-state)-this produces a $\delta^{(4)}(p_1+p_2-p_3)$ from the vertex integral. To see explicitly the interplay between presence of in/out states and time ordering, note that in a typical flat space in out correlator, everything including initial and final states is in time order; LSZ then converts it to an expression for $\langle f|i\rangle$, where a relative sign between momenta is induced because  $i\int d^4x e^{ipx}(\partial_\mu\partial^\mu+m^2)\phi=\sqrt{2\omega_p}(a_\infty-a_{-\infty})$.  This expression is, formally

\begin{align}
    |f\rangle=\sqrt{2k_1\cdot2k_2..}a^\dagger_{k_1}(\infty)a^\dagger_{k_2}(\infty)..|\Omega\rangle, |i\rangle=\sqrt{2p_1\cdot2p_2..}a^\dagger_{p_1}(-\infty)a^\dagger_{p_2}(-\infty)..|\Omega\rangle\\
    \langle f|i\rangle=\sqrt{2k_1\cdot 2k_2\cdot\cdot\cdot2p_12p_2..}\langle\Omega| a_{k_1}(\infty)a_{k_2}(\infty).....a^\dagger_{p_1}(-\infty)..|\Omega\rangle
\end{align}

And everything is naturally time ordered, and we use Feynman propagators to evaluate this expression via Wick's theorem. However, in an in-in correlator, we only have $|i\rangle$ state available, and therefore a correlation function is of the form $$\langle i| \mathcal{O}(t)|i\rangle\sim \langle 0|a_{p_1}(-\infty)a_{p_2}(-\infty)..\mathcal{O}(t)a^\dagger_{p_1}(-\infty)a^\dagger_{p_2}(-\infty)...|0\rangle$$

We must therefore resort to using Wightman functions, and this is precisely where destructive interference between modes can occur. For comparison, consider the difference between a sample in-out and in-in correlator. First,

\begin{multline}
   -i\lambda\delta(\sum\vec{k}_i)\int dt \langle k,out|T\phi_{k_1}(t)\phi_{k_2}(t)\phi_{k_3}(t)|p,p',in\rangle \\ \supset\langle a_k(\infty)\phi_{k_1}\rangle\langle \phi_{k_2} a^\dagger_p(-\infty)\rangle\langle \phi_{k_3} a^\dagger_{p'}(-\infty)\rangle\\
   \\=-i\lambda\delta(\sum\vec{k}_i)\int dt\delta(\vec{k}+\vec{k}_1)\delta(\vec{k}_2-\vec{p})\delta(\vec{k}_3-\vec{p'})e^{i(k-p-p')t}=-i\lambda\delta^{(4)}(k-p-p')
\end{multline}

Each contraction here is a Feynman propagator. The $1/2k$ normalizations cancel in the propagator and the definition of the external states.

To see what happens without an out state, let us compute the in-in correlator at equal-time insertions for this example. The first difference in the vertex integrals is the replacement

\begin{equation}
   \int_\infty ^\infty dt'\to 2Im \int^t_{-\infty}dt' 
\end{equation}

where $t$ is the time of insertion. With this replacement, the delta function for energies becomes a pole, with the expression(here $p$ is $p^0$) 
\begin{equation}
    \int_\infty ^\infty dt'e^{i(p_1+p_2+p_3)t'}=\delta(p_1+p_2+p_3)\to 2Im \int^t_{-\infty}dt' e^{i(p_1+p_2+p_3)t'}=\frac{2}{p_1+p_2+p_3}
\end{equation}

The question now is, in flat space, can we have physical poles like $(k_1+k_2-k_3)^{-1}$ in equal time in-in correlators. Clearly, these poles would correspond to an energy conserving delta function for a physical process $k_1,k_2\to k_3$. But these, as we saw, arise \textit{after} carrying out LSZ and placing some insertions in the far past and some in the far future-this cannot be done in an equal time in-in correlator. This was also observed in \cite{Green:2022fwg} where it was shown that LSZ killed all poles which didn't correspond to physical processes in flat space.

So then how do physical poles arise in inflation? First, note that in a general backgrounds the modes are complicated functions $u_k(t)e^{\pm(\omega t-\vec{k}\cdot\vec{x})}$. Secondly, in the in-in correlators, one replaces the Feynman propagator $D_F(t_1,t_2)$ with the \textit{Wightman function}, $G_W$. Recall that the flat space wightman function and feynman propagator are related as

\begin{align}
    D(t)=\langle 0|\phi_k(t)\phi_{k'}(0)|0\rangle=(2\pi)^3\delta(\vec{k}+\vec{k'})\frac{1}{2k}e^{-ikt}=D(-t)^*,\\
    D_F(t)=\theta(t)D(t)+\theta(-t)D(t)^*
\end{align}

We see that while the Feynman propagator contains both mode functions formally, it only ever has support on one of them-whichever is picked by time ordering. In contrast, the classical Green's function and the Wightman function for a Bugolyubov transformed mode is

\begin{align}
    G_c(k.t)=\frac{-i}{2k}(e^{ikt}+e^{-ikt})\\
    G_W^{BG}(0,\tau)=(2\pi)^3\delta(\vec{k}+\vec{k}')(Au_k(0)+Bu_k(0))(A^*u_k^* e^{ik\tau}+B^*u_ke^{-ik\tau})
\end{align}

where $u_k=\frac{H}{\sqrt{2k^3}}(1+ik\tau)$. Both of these contain both frequencies of modes, and it is natural to expect that they will generate a similar pole structure in correlation functions.  Note a number of things. Time ordering automatically ensures all these Wightman functions are just Feynman propagators, and the relative sign of the mode functions in the integral boils down to the existence of both in and out states.

To see this explicitly, suppose we instead calculated the (physically nonsensical) time ordered quantity $\langle 0|\phi_1\phi_2\phi_3|k,p,p',in\rangle$ where there's no out state. Then, the first (time ordered) contraction above is replaced by $\langle a_k(\infty)\phi_{k_1}(t)\rangle\sim e^{ikt}\to\langle \phi_{k_1}(t)a^\dagger_k(-\infty)\rangle \sim e^{-ikt}$, where the physical meaning changes from creation at the vertex to annihilation. This new correlator evaluates to $-i\lambda \delta^{(4)}(k+p+p')$, which ofcourse cannot be physically reached, but has the same energy-momentum dependence as the Bunch Davies in-in correlator where we would see a $1/(k+p+p')$ pole for mode insertions with momenta $k,p,p'$. We see clearly the role of the presence of an out-state in reaching physical singularities, and the presence of unphysical total energy singularities in the presence of only in states. The relationship between time ordering and the presence of out states can also be seen by noting that $\langle a_k(\infty)\phi(t)\rangle$ doesn't vanish because of how time ordering places the $a$ so that it acts on the bra. If we had an in-state on the left, this would be replaced by $\langle a_k(-\infty)\phi(t)\rangle$ which would be killed by time ordering as it would take $a$ to the right and annihilate the vacuum. Thankfully, this such a correlator wouldn't be time ordered anyway and therefore we'd be forced to use Wightman functions instead of the Feynman propagator, and they don't vanish.

Now, consider a correlation function with a highly occupied state $N$, of the form $\int d^4x \langle N|T \phi\phi\phi|N\rangle$. By construction, this correlator is a physical process with external states put on shell. Therefore, integrating over the vertex will produce physical poles depending on which mode of the  initial state we use to contract with the operators inside the correlator. We can pick out 2 $k_i's$ from the in state and one from the out state and get a delta function that conserves the in-out momentum, which in the in-in case would produce a physical pole. In the language of flat space Feynman rules, these physical delta functions arise because of the fact that contractions with in and out states induce exponentials with different signs of momenta-indeed this is due to the implicit LSZ that has been carried out already where we know that the exponentials for in/out states are hermitian conjugates of each other. 

%Just to emphasize this point, note that in flat space scattering, the final and initial states were defined to be annihilated by $a(\pm\infty)$ respectively. Then, one used the identity $i\int d^4x e^{ipx}(\partial_\mu\partial^\mu+m^2)\phi=\sqrt{2\omega_p}(a_\infty-a_{-\infty})$ and its hermitian conjugates for the $a^\dagger$. Therefore, the fact that we had states of the type $\langle f|=\langle \Omega|a^\dagger(\infty);|i\rangle=a(-\infty)|\Omega\rangle$ automatically induced a relative sign between their momenta in the exponential prefactor for LSZ, and therefore gave rise to physical delta functions. 

This is the state of affairs. In the presence of particles in the external states, the possibility of physical pole arises because they may either annihilate or be created at a vertex. In an in-out correlator, the relative sign between the time of insertions for in and out states(which is in practice consistently accounted for via LSZ) between these processes reflects in physical singularities. In an in-on correlator, operationally, this reflects in the fact that $\langle a \phi_p(t)\rangle\sim e^{ikt}$ whereas $\langle \phi_p(t)a^\dagger\rangle\sim e^{-ikt}$. We will see that physical poles will come from contractions schematically of the type

\begin{equation}
    \langle ...p..|\zeta_1\zeta_2\zeta_3|...p..\rangle\supset\langle p\zeta_1\rangle\langle \zeta_2\zeta_3p\rangle 
\end{equation}

where we indicate the terms which contract together at a vertex(the second term), and the term that spectates(the first term).
 In the absence of particles in the initial state, or the inability to perform LSZ; correlation functions will only depend on the Wightman functions between field insertions, and in time dependent backgrounds it is possible-in fact necessary- to have Wightman functions that include both modes(as opposed to the Feynman propagator which picks only one based on time ordering) and therefore produce physical singularities through their interference terms, despite the absence of features like time ordering and LSZ in in-in correlators. In curved space, we will use the trick of performing calculations in a frame where there are no particles, but the Wightman functions include both modes and consequently help achieve the same physical singularities as one would expect in the presence of particles. 

The occurrence of physical poles in cases where this can't be done and therefore initial state particles interacting at a vertex cannot be avoided is discussed in \ref{sec:classical}. With particles in the initial state, an in-in correlator receives contributions from terms where one of the initial state particles is contracted with the (cubic)vertex insertion along with 2 of the field insertion, while the third spectates. This contraction generates a physical pole at the right momentum support for the spectator. In section \ref{oneparticleBD} the occurrence of this pole in the presence of a single particle in the initial state was studied in detail.

 \section{Choice of interaction and their imprint on singularities}\label{sec:interactions}

The role of singularities of the S-matrix in flat space has a clear physical meaning and origin. The role of poles and branch cuts are easy to interpret from their physical meaning, which lies in the Kallen-Lehmann decomposition of correlators. The nature of singularities is determined by the nature of interactions considered and the kind of processes they involve(for instance, tree level exchanges generates simple poles whereas loops are generally accompanied by branch cuts). Correlation functions are expected to be constrained by Lorentz invariance, and in flat space for example we have delta function singularities indicating Poincare invariance that conserves momenta. In deSitter, the correlation functions are similarly constrained by the deSitter isometries which are conveniently those of an Euclidean CFT \cite{Creminelli:2011mw,Kehagias:2012pd,Antoniadis:2011ib}. We will, for completeness, briefly discuss the nature of singularities with different interactions in deSitter. The point is to emphasize that \textit{while the functional form of singularities can change depending on the choice of interaction, the object we are trying to probe is the momentum structure encoded in the singularities, which carries a clear physical signature of the choice of quantization.} Let us begin by noting that far outside the horizon, the behaviour of a scalar field $\phi$ that has a very small coupling to the  inflaton is of the form

\begin{equation}
    \phi\sim \tau^\Delta\hspace{5mm}\Delta\equiv\frac{3}{2}\bigg(1-\sqrt{1-\frac{4m^2}{9H^2}}\bigg)
\end{equation}

In this section we will focus on the case of Bunch Davies modes in a Bunch  Davies initial state. Next, note that the correlators must be functions of the de Sitter invariant distance-

\begin{equation}
    d(x_i,x_j)=\frac{|\vec{x}_i-\vec{x}_j|^2}{\tau_i\tau_j}-\bigg(\frac{\tau_i}{\tau_j}+\frac{\tau_j}{\tau_i}\bigg)\to \frac{|\vec{x}_i-\vec{x}_j|^2}{\tau_i\tau_j}|_{\tau\to 0}
\end{equation}

Note that, in particular, this means that the spatial dependence of the correlation function is specified automatically once the time dependence of the field(which is related to its mass) is specified at late times \cite{Creminelli:2011mw}. The Fourier transform of this correlator contains the momentum dependence we are interested in, and this tells us that once the late time behaviour of the field is specified, the momentum dependence is specified too. Let us see how this works with the simple example of an interaction $\sqrt{-g}\lambda\phi^3/6$. For a general $\Delta$, the time dependence of the 3 point function will be fixed as $\sim\tau_1^\Delta \tau_2^\Delta \tau_3^\Delta$, and therefore a generic  position space function can only take the form

\begin{equation}
    \langle \phi_1\phi_2\phi_3\rangle\sim \prod\frac{\tau_i^\Delta}{|\vec{x}_i-\vec{x}_j|^\Delta}
\end{equation}

Let us see what happens for different values of $\Delta$. For a field that has mass $m=\sqrt{2}H$ we have a conformally coupled scalar with $\Delta=1$. The Fourier transform of $\prod 1/|\vec{x}_i-\vec{x}_j|$ goes as $1/(k_1k_2k_3)$, and indeed in momentum space we find with a full in-in calculation,  upto a 3-momentum conserving delta function

\begin{equation}
    \langle \phi_{k_1}\phi_{k_2}\phi_{k_3}\rangle=\frac{\pi\lambda H^2}{8}\tau_*^3\frac{1}{k_1k_2k_3}
\end{equation}

where $\tau_i=\tau_*$ is imposed for an equal time correlator. Note that the result is dependent on $\tau_*$. To study the limit $\Delta\to 0$, which is more subtle because now the late time behaviour of fields isn't a simple power law, one can use the fact that $\lim_{\Delta\to 0}\tau^\Delta\sim 1+\Delta\log\tau$ and anticipate a logarithmic dependence on the late time, and therefore, on momenta. This is to be expected, since the field has no mass of it's own and therefore the only kinematic quantity available to make the logarithm dimensionless is a linear combination of momenta. The classical scale invariance of the massless theory is broken when one imposes a regulator(in the form of a finite cutoff $\tau_*$), and this breaking is captured in the logarithmic piece. Maldacena's consistency requirement directly relates this nongaussianity to a breaking of scale invariance, as we shall later see. It is important to note that the massless case is subtle, because the limits $\tau\to0,\Delta\to 0$ don't commute, and a naively $\tau$-independent field outside the horizon acquires a logarithmic time dependence due to the cubic interaction. This logarithmic dependence on time then fixes the momentum dependence because of the nature of dS invariant distances. The in-in calculation in momentum space yields, as expected 

\begin{equation}
    \langle \phi_{k_1}\phi_{k_2}\phi_{k_3}\rangle=\frac{H^2\lambda}{12(k_1k_2k_3)^3}\bigg(k_1k_2k_3-\sum k_ik_j^2+\sum k_i^3(-1+\gamma+\log(k_t\tau_*))\bigg)
\end{equation}

This is where the role of derivative interactions becomes important. For a massless field, the apparent singularity at $\tau_*\to 0$ can be avoided by considering only derivative interactions. The mathematical origin is simple-the interaction hamiltonian in conformal time is of the form $\int \frac{\lambda}{6\tau^4H^4}\phi^3$ which is ill defined at $\tau=0$. In the above 3 point correlator, the origin of the logarithm is from integrals of the form $\int e^{ik_t\tau}/\tau$ which are singular at the point $\tau=0$. If we had sufficient powers of $\tau$ in the numerator, then this problem can be cured. With a purely cubic interaction, the Wightman functions entering the in-in formula are $\int \frac{d\tau}{\tau^4}\prod_i\langle \phi_i(0)\phi(\tau)\rangle\sim\int \prod_i(\frac{1}{2k_i^3}(1-ik_i\tau)e^{ik_i\tau})$ where $\tau$ is at the vertex being integrated over in the interaction hamiltonian. If, however, we consider the Wightman function for $\langle\zeta\zeta'\rangle\sim \sqrt{k}\tau e^{ik\tau}$, then we see that the numerator's factors of $\tau$ can compensate  for the ill defined $\tau\to 0$ behaviour of the Hamiltonian. With an interaction like $\zeta'^3$, the interaction hamiltonian is $\int \frac{\lambda}{6H\tau}\zeta'^3$, and the late time divergence is softer. Then, the in in calculation will contain integrals of the type $\int \frac{d\tau}{\tau}\prod_i\langle \phi_i(0)\phi(\tau)\rangle\sim\int d\tau \tau^2\prod_ie^{ik_i\tau}$, which has no log divergences. The singularities are in fact simple poles of the form $1/k_t$. In fact, it suffices to have an interaction of the type $\zeta\zeta'^2$-this is indeed a leading cubic interaction in Hamiltonian, where the $1/\tau^2$ in the denominator of the interaction hamiltonian can get compensated by two powers of $\tau$ coming from $\zeta\zeta'$ contractions; the remaining $\zeta\zeta$ contraction is free of any $1/\tau$ divergences.

Thus, we see that considering derivative interactions softens the behaviour of the $\tau\to 0$ limit of the correlators, and one needn't conduct any explicit regularization. We see that one needs \textit{a minimum of two derivatives} in the interaction to get rid of this regulator requirement, and for simplicity we consider a $\zeta'^3$ interaction.

Now, the crucial point we wish to emphasize. While the nature of singularities might be different in their functional forms (logarithmic, simple poles, or a combination); the momentum dependence of these singularities for the Bunch Davies case is always $(k_1+k_2+k_3)$. What we aim to understand is-having fixed an interaction and therefore the functional form of the singularity- does the nature of this momentum dependence change once we allow different initial states and/or mode functions? The answer, as we see, is quite generically yes.

 \section{Comments on the Maldacena consistency criteria}\label{sec:Maldacena}
 
 The Maldacena consistency criteria is a consequence of frequency-dependent freezing times of various modes. Consider the three point function in the squeezed limit $k_1<<k_2,k_3$. In this case, the mode $\zeta_{k_1}$ freezes in (becomes a classical variable) before the modes $\zeta_{k_2,k_3}$ have crossed the horizon. It then serves as a background over which the latter modes propagate. In effect, the 3 point function reduces to 
 
 \begin{equation}
     \langle \zeta_2\zeta_3\rangle_{\zeta_1}\approx\langle \zeta_1\zeta_2\rangle_0+\zeta_1\frac{\partial}{\partial_{\zeta_1}}\langle \zeta_1\zeta_2\rangle_0
 \end{equation}
 
 The fact that $\zeta_1$ has frozen in can be accounted for by a rescaling of coordinates-in particular-
 
 \begin{equation}
     ds^2=dt^2-a^2e^{2\zeta_1}(dx^2+dy^2+dz^2)\implies x\to (1+\zeta_1)x
 \end{equation}
 
 Thus, the squeezed limit can be see to be accounted for by spatial reparametrizations. Then, in position space one can account for invariance under spatial reparametrizations as
 
 \begin{equation}
     \langle \zeta(x)\zeta(0)\rangle_{\zeta_1}=\langle\zeta(x+\zeta_1x)\zeta(0)\rangle_0
 \end{equation}
 
 We can now Taylor expand the RHS too, to get a term of the form $\zeta_1x\frac{d}{dx}\langle \zeta\zeta\rangle$ and then multiply both sides with $\zeta_1$ before averaging. Converting to momentum space, this yields
 
 \begin{equation}
     \lim_{k_1<<k_2,k_3}\langle \zeta_1\zeta_2\zeta_3\rangle\approx\langle\zeta_{k_1}\zeta_{-k_1}\rangle k\frac{d}{dk}\langle\zeta_{k_2}\zeta_{k_3} \rangle
 \end{equation}
 
 Now, for a \textit{nearly scale invariant power spectrum}, $\langle \zeta\zeta\rangle\sim k^{-3+n_s}$, and therefore we derive a consistency equation
 
 \begin{equation}
     \lim_{k_1<<k_2,k_3}\langle \zeta_1\zeta_2\zeta_3\rangle=-n_s\langle \zeta_{k_1}\zeta_{-k_1}\rangle\langle\zeta_{k_2}\zeta_{k_3}\rangle
 \end{equation}
 
 Note the crucial step where we absorbed the background field into the metric, therefore finding ourselves a coordinate system where the power spectrum is insensitive to the background wave\cite{Creminelli:2004yq}. This step was valid because at leading order, $\zeta$ interactions come from the nonlinearities of the metric itself, without any reference to the features of the potential. However, the interaction we have considered comes  from a potential with derivative interactions $\dot{\phi}^4$. It is known \cite{Creminelli:2003iq} that nongaussianities from such interactions are suppressed by $\frac{k_L^2}{k_S^2}$ in the squeezed  limit, where $S,L$ are the large and small momenta respectively( the subscripts are conventionally chosen to denote short and large distances). The basic idea is that in the squeezed limit, the relevant interactions are those where one of the insertions has no derivatives, and this is the insertion that's chosen to be the short wavelength mode. Explicitly, the relevant action at cubic order is \cite{Creminelli:2011rh}
 
 \begin{equation}
     S=\int d^4x \epsilon a^3\bigg((1+3\bar{\zeta})\zeta'^2-(1+\bar{\zeta})(\partial_i\zeta)^2\bigg)
 \end{equation}
 
where the insertion that's taken to be the squeezed wavelength has been made explicit. Any corrections to this will therefore be from derivative corrections, and it was argued in \cite{Creminelli:2011rh} that the corrections are atleast quadratic in $(k_L/k_S)$. There are many ways to see this, starting from nothing that the power spectrum at late time recieves genuine corrections only from second derivatives, because $\langle \zeta_1\zeta_2\rangle=P(k_1)(1+k^2\tau^2)$ and therefore if one was to organise this expression as a power series in $k\tau$, the leading correction at $\tau\to 0$ would come from $\partial_\tau^2$, not before. Yet another way to see this is to note that schematically, the Wightman functions $\langle \zeta\zeta\rangle$ and $\langle\zeta\zeta'\rangle$ differ by a factor of $k^2\tau$ and therefore in the in-in formalism, we replace the former with the latter(as is what  happens if we switch from inflation's leading interaction to our), we get additional powers of $(k_L/k_S)^2$. On more general grounds, focusing on the momentum dependence and ignoring the prefactors, the bispectrum is always of the form 

\begin{equation}
    \langle \zeta_1\zeta_2\zeta_3\rangle\sim \frac{1}{(k_1k_2k_3)^3}f(k)
\end{equation}

And in the squeezed limit where we denote $k_1\to k_L,k_{2,3}\to k_S$, we find that in the case where $\zeta$ is only a (scalar) mode of the metric, $f(k)\to k_L^3$ therefore restoring the consistency criteria. However, with an interaction like ours, $f(k)$ includes term of the nature $k_L^2k_S$ because now the short wavelength can be put onto any of the three insertions in the interaction hamiltonian without being ignored unless we wish to kill the entire contribution. This produces the required $(k_L/k_S)^2$ suppression. As has been emphasized in \cite{Creminelli:2011rh}, $f(k)\to 0$ as $k_L^2$, not $k_L$. On the other hand, without higher derivative features in the potential,  $f(k)$ is a constant in $k_L$. For an explicit example, consider our eq(\ref{eq:3pt}), where the bispectrum is

\begin{equation}
    \Delta\langle \zeta_1\zeta_2\zeta_3\rangle\sim \frac{1}{(k_1k_2k_3)(k_1+k_2+k_3)^3}\to P(k_s)P(k_L)\bigg(\frac{k_L}{k_S}\bigg)^2
\end{equation}

And therefore we see that the consistency condition is violated by an amount, as has been argued on general grounds,  $\left(1+\bigg(\frac{k_L}{k_S}\bigg)^2\right)$.

 The behaviour of $f(k)$ is more interesting in the case we have physical poles. The general conclusion is that this time, the behaviour is modified by a reciprocal quantity-$k_S/k_L$ which is severely enhanced in the squeezed limit. In the squeezed limit, for all cases other than choosing purely BD initial conditions, the consistency criteria fails. The reason is the presence of physical poles, which generate terms proportional to $k_1/k_3$ which swamp the $\vec{k}_3\to 0$ limit and spoil the consistency. The spoiling depends on the angle between the momenta, because we have a constraint $\sum\vec{k}_i=0$ from which we can study the poles in terms of the magnitudes $k_i$ as well as the angles. It is clear that such a structure wouldn't occur if we had only physical poles, because then all 3 momenta are on an equal footing inside the correlator and setting any one to zero can only generate $k_1+k_2$ poles, not a $k_1-k_2$ pole that can potentially compete with the $k_3\to 0$ pole and therefore require to be kept intact before taking the limit. Since the consistency condititon is linked to spatial reparametrization invariance, it is tempting to think that OPE, gauge choice and the BD initial conditions are all intimately linked. In fact, it was argued in \cite{Shukla:2016bnu} how the reparametrization invariance is a feature of Einstein's equations, and with a non-BD vacuum, we must account for the effect of it's contributions to the stress energy tensor's vacuum expectation value. Neglecting this contribution means that we have not fully accounted for the modification to the ADM action-the piece which actually solves Einstein's equations. Therefore, we are working with an action that doesn't yield the correct equations of motion, and therefore we have lost the freedom to carry out spatial reparametrizations, and consequently the consistency condition. 
 
 We therefore see that while the squeezed limit doesn't directly contain information about the analytic structure of the bispectrum because the modes have been in a sense fixed; it does carry information about the initial states and the vacuum of the modes in the form of departures from Maldacena consistency. In particular, a $(k_L/k_S)^2$ departure is a signature of a Bunch Davies initial state with higher order interactions, whereas everything else produces a large deviation in the form of $k_S/k_L$ poles.
 
 \section{Classical vs Quantum}\label{sec:classical}
 
The authors of \cite{Green:2020whw,Green:2022fwg} argued that physical poles are a reflection of a classical initial state. At the level of calculation, it amounted to using the classical Green's function, 
 
 \begin{equation}
     G_{c}(0,t)=\frac{-i}{2k}(e^{ik\tau}-e^{-ik\tau})
 \end{equation}
 
 which contains both positive and negative frequency modes, and therefore the product of these Green's functions generates poles with varying momentum dependence depending on the interference terms between the modes. However, the Wightman function for purely Bunch Davies modes contains only one of these modes, whereas the one for Bugolyubov transformed states(for example the alpha vacuum) contains both-
 
 \begin{equation}
     G_W^{BD}(0,\tau)\sim u_k(0)u_k^*(\tau)e^{ik\tau},G_W^{\alpha}(0,\tau)\sim v_k(0)v_k^*(\tau)=v_k(0)(A^*u_k^*(\tau)e^{ik\tau}+B^*u_k(\tau)e^{-ik\tau})
 \end{equation}
 
 Therefore, we clearly see the link between the appearance of mode mixing in the case of classical as well as Bogolyubov transformed states, and the subsequent appearance of physical poles. What about the case of excited states? The presence of particles in the initial state means that there are additional possible interactions at a vertex, where some of the field insertions annihilate not just each other but also these initial state particles at a vertex. This is drastically different from the case of a vacuum initial state where only the field insertions annihilate(or create) among themselves at a vertex, and the only kinematic constraint then is conservation of their spatial momenta. The next crucial difference is the interplay of energies(stacked in the exponentials in the mode functions) at the vertex. In the cases where the initial state particles don't interact at a vertex but merely spectate, the vertex represents a \textit{creation} of particles from the vacuum, as in the figure (cite figure). Since the vertex factor for creation at a vertex is accompanied by the mode $e^{ik\tau}$, the creation of three particles comes with a $\tau-$integral over $e^{i(k_1+k_2+k_3)\tau}$, and produces a total $k$ pole as we have seen. However, the vertex factor that accompanies an annihilation is quite different, and has the opposite energy as the mode function for creation. If a particle in the initial state with momentum $P$ annihilated at a vertex to produce 2 particles of momenta $k_1,k_2$; then we would find the prefactor at the vertex to be $e^{i(k_1+k_2-P)\tau}$, along with a momentum conserving delta function for the remaining particles which is just an exchange of momentum, and not a genuine interaction. Together, they produce a physical pole, and it's origin is clear-it's the relative sign of momenta in the mode functions corresponding to annihilation vs creation. However, because of these same momentum conserving delta functions, these physical poles have a very restricted support and will therefore not be observable except at isolated points in phase space. 
 
\begin{figure}
     \centering
     \includegraphics{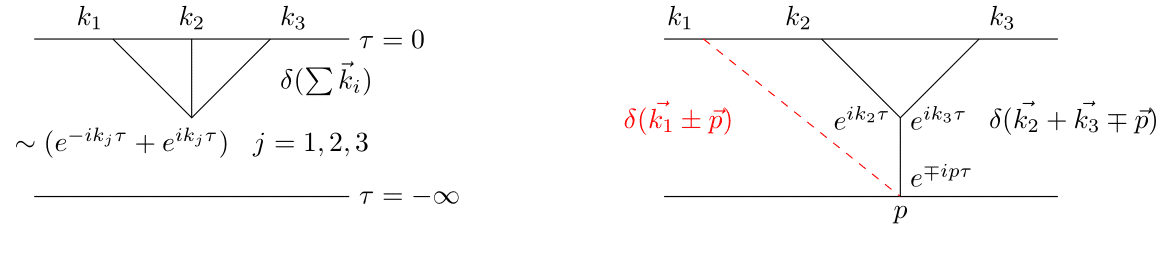}
     \caption{(Left) the case of classical or accelerated observer's mixed mode functions, (Right) the case of genuine interaction of initial state particle. The $\pm$ sign is determined by whether the particle is created/annihilated at the vertex. The latter generates physical poles. The red dashed line represents a spectator.}
     \label{fig:1}
 \end{figure}

 An obvious way to get rid of the isolated  nature of these singularities is to average over all possible momenta of initial state particles-this smears out the phase space where the poles have support. Crucially, \textit{this cannot be accomplished with a state where a only fixed number of momentum states are occupied, no matter how highly.} That is to say, states of the type $\otimes_k (a^\dagger_k)^n|0\rangle$ cannot display observable physical poles unless we consider a sum over all possible momenta at each order in $n$. Schematically, a correlator like $\langle P| \zeta_1\zeta_2\zeta_3|P\rangle$  won't generate observable physical poles, but a correlator like $\int d^3P \langle P| \zeta_1\zeta_2\zeta_3|P\rangle$ will. Indeed, this is what we saw for states excited with respect to the Bunch Davies vacuum. As we introduce more and more particles, we must accept the possibility of more and more possible disconnected pieces and consequently isolated poles at a given order in perturbation theory, and it seems unlikely that we will ever be rid of this problem.  However, for \textit{some} excited states (relative to a Bunch Davies observer), it is possible to go to a frame where this state is a vacuum, and there are no particles in the initial state to interact at a vertex. These states are precisely the Bugolyubov transforms of the Bunch Davies vacuum, which are excited for a Bunch Davies observer but have zero net momentum. If we compute these correlators in a frame where the particles actually exist, we see(\ref{sec:alpha}) indeed the presence of physical poles corresponding to physical interactions of initial state particles at a vertex, but it is difficult to find a clear interpretation due to additional disconnected pieces that must exist for reasons mentioned above. Instead, calculating them using the accelerated observer's basis ensures that these particles are not seen, but their physics is now captured in the modified mode functions-specifically, we must now also include negative frequency modes that otherwise appear due to annihilation of particles at a vertex. The physics of annihilation at a vertex is now absorbed into the physics of the new mode functions, and the mathematical correspondence is clear. Figure \ref{fig:1} demonstrates this discussion. (Here, by 'create/annihilate' we simply mean the way in which they contract-either with an $a_p$ or with $a^\dagger_p$).
 
 To contrast the situation with flat space time ordered in-out correlators, recall that the frequency of the mode functions there is fixed via LSZ, and time ordering picks the right frequencies in the Feynman propagators. In contrast, for in-in correlators, both the classical Green's function and the quantum Wightman functions for Bugolyubov states contain both signs of frequencies. Even in flat space, the Wightman function can have either frequency depending on the time order of the field insertions $\langle \phi_k(t)\phi_{k'}(0)\rangle\sim \delta(\vec{k}+\vec{k'})e^{-ipt}/2k=\langle \phi_k(0)\phi_{k'}(t)\rangle^*$; where the complex conjugation is equivalent to a negative frequency. It is precisely time ordering that eventually tells us which frequency to choose. The lack of this luxury with in-in correlators is what is ultimately tied to physical poles.
 
 \section{Choosing a cutoff} \label{sec:cutoff}

 While carrying out the integrals  in the in-in correlators, the lower limit $\tau\to -\infty$ must be taken with care \cite{Chen_2007,Chen_2010}. Generally, the exponential factor at the lower limit is taken to vanish because of it's oscillatory nature \cite{Ganc:2011dy}. The general idea is that \cite{Holman:2007na} the initial conditions need to be set when the EFT of inflation can still be trusted, and the Bunch Davies prescription is singled out by setting them at $\tau\to -\infty$. However, assuming that there is some cutoff scale $\Lambda$ upto which our EFT can be trusted, then the validity of this EFT is reached at some $k\sim a(\tau_0)\Lambda$ where $k$ is a physical momentum scale today-note that inflation redshifts energies. Scales that were once in the UV become part of the IR due to the universe's expansion. Setting a cutoff at a finite $\tau_0$ automatically takes us away from the Bunch Davies prescription, and we need to introduce more general initial conditions, which yields modes which are Bogolyubov transformations of Bunch Davies. We would like to understand if taking the lower limit $\tau\to -\infty$ in our calculations hides any information about the physical poles seen by an accelerating observer-in particular, if the singular behaviour at physical energies is still seen. If we instead keep a cutoff intact, then we see that the integrals containing physical poles are modified to($k'_j\equiv k_t-2k_j$)
 
 \begin{multline}
     \textrm{Im} \int_{\tau_0}^0 d\tau \tau^2 e^{ik'_j\tau}=\textrm{Im}\frac{i(2+e^{ik'_j\tau_0}({k'}_j^2\tau_0^2+2ik'_j\tau_0-2))}{{k'}_j^3}\hspace{3mm}\textrm{(with $\zeta'^3$ interaction)}\\\label{7.1}
     \\=\frac{2}{{k'}_j^3}(1-\cos(k'_j\tau_0))+\frac{\tau_0^2 \cos(k'_j\tau_0)}{k'_j}-2\frac{\tau_0\sin(k'_j\tau_0)}{{k'}_j^2}
 \end{multline}
 
 We need to check the leading singular behaviour as $k'_j\to 0$. Expanding the numerators for small $k'_j$(but finite $\tau_0$) we find
 
 \begin{equation}
     \textrm{Im} \int_{\tau_0}^0 d\tau \tau^2 e^{ik'_j\tau}|_{k'_j\to 0}=\frac{2\tau_0^2}{k'_j}-\frac{2\tau_0^2}{k'_j}+\textrm{terms regular in $(k'_j\tau_0)$}
 \end{equation}
 
 From this expression, it is clear that the limits $k'_j\to 0$ and $\tau_0\to-\infty$ don't commute, and in fact taking $\tau_0\to-\infty$ before going to the folded configuration $k'_j\to 0$ displays a divergence(from the terms regular in $\tau_0$ which blow up in this limit), whereas the opposite order of limits is finite. This transparently demonstrates that the divergence in the folded limit is an artifact of the initial conditions being set at $\tau_0=-\infty$. The absence of a divergence can directly be seen by plotting \eqref{7.1} as a function of $k'_{j}$.
\begin{figure}[h]
\centering
\includegraphics[width=12cm, height=8cm]{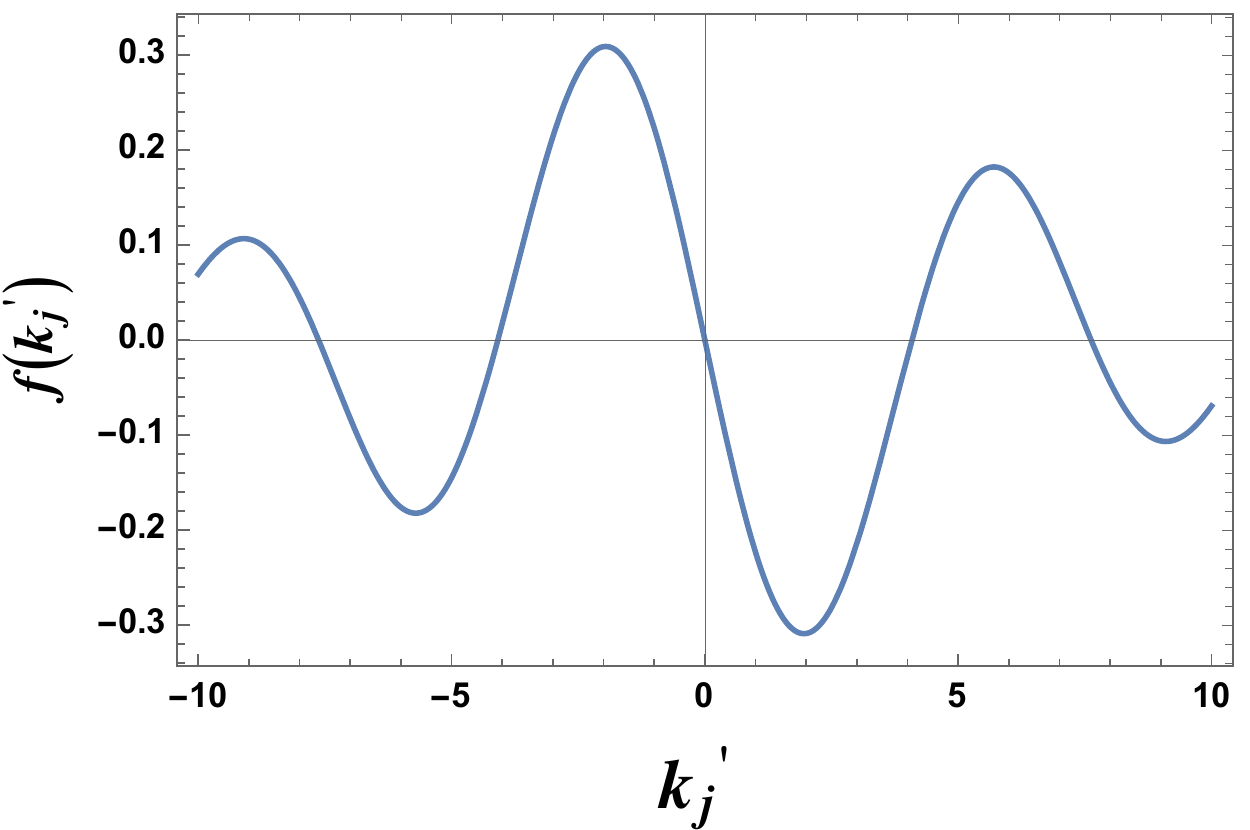}
\caption{Plot of $f(k'_j)$(see \eqref{7.1}) as a function of $k'_{j}$. This shows that there is no singularity in the folded limit. In addition to this there is an oscillatory feature.}
\label{fig:cutoff}
\end{figure}
 Therefore, clearly there is no singularity in the folded limit. This can be seen directly from Fig~\ref{fig:cutoff} where we have plotted \eqref{7.1} as a function of $k'_{j}$.
 It has been noted earlier too, for example in \cite{Chen_2007}, that the divergence in the folded limit can be cured by introducing an early time cutoff. In \cite{Holman:2007na}, a similar phenomenon was shown for the interaction of inflation, where the presence of a cutoff rendered the folded limit finite but regular in the cutoff, and would therefore be subject to the same conclusion as ours. For an initial state that is a Bogolyubov transform of the Bunch Davies vacuum, setting the initial conditions away from $-\infty$ is perhaps  reasonable because one can argue that a Bogolyubov transformed operator basis at $\tau_0$ can be thought to be time evolved from a Bunch Davies basis which was imposed at $\tau\to-\infty$ . However, there is no first-principles argument for this, and there are other excited states where this reasoning wouldn't be straightforward. 
 
 In the literature, one will find plentiful examples of both cases - where initial conditions are either set at $-\infty$ or at a finite cutoff. Since the primary motive of our analysis is to study and compare the pole structure of correlators, and trace the mathematical origin of physical poles in quantum and classical correlators, we will not dwell on this issue further, and treat our initial conditions set at $-\infty$.
 
 \section{Conclusions}
 
 Our results can be summarized as follows (BD: Bunch Davies, BG-Bogolyubov transform)
 
 \begin{center}
     \begin{tabular}{c|c}
         Initial state & Pole structure  \\
         \hline
          BD vacuum & $k_t$ pole\\
          Excited over BD with finite modes present & disconnected physical poles, connected $k_t$ pole\\
          Excited over BD with all modes present & connected $k_t$ pole, observable physical pole\\
          Excited over BD but a BG & both physical and $k_t$ poles\\
          Excited over BD but coherent & $k_t$ pole\\
          Excited over BD but coherent for some BG & both $k_t$ and physical poles\\
          Classical & both $k_t$ and physical poles
     \end{tabular}
 \end{center}

 We find that despite the rich variety of possible initial states and vacua choice for the modes, the analytic structures of correlators - in particular the bispectrum - is largely of two types. The case where the initial state is Bunch Davies has a characteristic total energy pole, whereas all other possibilities exhibit in addition physical poles. An interesting exception is that of Coherent states (under the assumption of vanishing one point function), which are excited states but when built with Bunch Davies particles, don't display any physical poles. These poles are restored for Coherent states relative to Bugolyubov transformed observers. In fact, it makes more sense to note that vacua are just special(trivial) cases of coherent states, and therefore the pole structure obtained with a coherent state is expected to be reproduced for vacua. Departures from the Bunch-Davies pole structure are also encoded in the sharp $k_{S}/k_{L}$ violation of the Maldacena consistency criteria(where $k_L$ is the small momentum) in the presence of physical poles, and therefore the squeezed limit of the bispectrum is a useful observational tool to characterise the initial state. It is tempting to interpret these as physical scattering/decay processes in the initial state, as was hinted in \cite{Green:2020whw}; but we show that this interpretation is unclear. For one, coherent states with vanishing one point functions don't display any physical poles, despite containing excitations. On the  other hand, an alpha vacuum that serves as the initial state for the modes as well as inflation exhibits physical poles which cannot be ascribed to alpha-particles. One could in principle circumvent this apparent discrepancy by demanding that by particles one means only BD particles, but this is against the spirit of general covariance and the well known fact that such an interpretation is not unique. We see that initial state effects on the bispectrum lead to enhancements in certain momentum configurations; a fact that has been noted before in the literature. We have attempted to unearth the connection between the mathematical machinery that leads to physical poles  in a classical calculation vs an in-in calculation, as well as the corresponding intuition from flat space. We have also made clear the role of derivative interactions in generating singularities which are simple poles. We believe making these associations transparent clarifies much of our understanding about correlation functions in deSitter and the origin of their singularities.
 
 \section*{Acknowledgments}
 
 We would like to thank Aleksandr Azatov, Paolo Creminelli, and Sachin Jain for valuable feedback and discussions. We thank Sounak Sinha for initial collaboration in the project. DG acknowledges support through the Ramanujan Fellowship and MATRICS Grant of the Department of Science and Technology, Government of India.

\appendix  
  \section*{Appendix}

  \section{Perturbations and perturbation theory}
The inflationary background is a quasi-de Sitter spacetime with deviations from pure Sitter being slow-roll suppressed. In this paper we consider a pure de Sitter background and analyse correlations of the adiabatic fluctuations $\zeta$ about the background. Before embarking on calculating correlation functions of the perturbations, It is essential to clarify the meaning of perturbations in a general curved background. Since, we are free to perform arbitrary coordinate transformations we can remove or introduce fictitious perturbations which makes cosmological perturbation theory subtle\cite{Bardeen:1980kt}. Consider a homogeneous function $\rho(t)$ which has no spatial dependence. Now, perform the following coordinate transformation
\begin{equation}\label{4.1}
    x^{\mu}\rightarrow x'^{\mu}=x^{\mu}+\epsilon^{\mu}(x)
\end{equation}
Clearly in the primed coordinate system the function $\rho$ now also has spatial dependence. The fact is that one has to take into account full set of perturbations i.e field perturbations + metric perturbations and one can push the perturbations entirely into the metric or the field configuration but cannot get rid of them completely. \par
The field/metric perturbations are defined as follows
\begin{equation}
    \delta Q(x)=Q(x)-\overline{Q(t)}
\end{equation}
where $Q(x)$ is any function in real-physical spacetime and $\overline{Q(t)}$ is a background function we associate to $Q(x)$. This is the so called passive approach to perturbation theory as described in\cite{Mukhanov:1990me}. Here, $Q$ can represent a scalar, vector  or even a tensor quantity. The background function is supposed to represent the evolution of the unperturbed homogeneous universe and is fixed i.e after a coordinate transformation $\overline{Q(t)}\rightarrow \overline{Q(t')}$, the function does not change!
Since, the background is time-dependent even the scalar perturbations are not gauge-invariant. For concreteness let us work out transformation law for scalar perturbations. 
\begin{equation}
    \delta Q(x)=Q(x)-\overline{Q(t)}
\end{equation}
Now perform the infinitesimal transformation given in \eqref{4.1} and we have 
\begin{equation}
   \delta Q'(x')=Q'(x')-\overline{Q(t')}
\end{equation}
Notice, that the background function does not change as mentioned above. On the RHS we can replace $Q'(x')$ by $Q(x)$ since it is a scalar function and we have
\begin{align}
     \delta Q'(x')=Q(x)-\overline{Q(t)}-\epsilon^{0}\dot{\overline{Q(t)}}
\end{align}
The dot represents time derivative. Now, we can combine the first and the second term on the RHS giving the perturbation in the unprimed coordinates and we get our final expression
\begin{equation}
     \delta Q'(x')=\delta Q(x)-\epsilon^{0}\dot{\overline{Q(t)}}
\end{equation}
It is very clear (as mentioned above) that if the background is not time dependent then the perturbations are gauge invariant.\par 
Since, the perturbations are gauge dependent, It is useful to work with Gauge-Invariant variables instead. One of most used variable is the co-moving curvature perturbation denoted by $\zeta$, which equals the curvature perturbation in the co-moving gauge defined below.
\begin{equation}
\delta\phi=0, \hspace{20mm} g_{ij}=a^{2}\left(\left(1+2\zeta\right)\eta_{ij}+\gamma_{ij}\right),
\hspace{20mm}\partial_{i}\gamma_{ij},
\hspace{20mm}\gamma_{ii}=0
\end{equation}
where $a$ is the scale factor. Notice, in this gauge the scalar perturbation is completely absorbed in the metric as $\zeta$, defined as
\begin{equation}
\label{7.2}
    \zeta=\Psi+H\frac{\delta\phi}{\dot{\phi}}
\end{equation}
by construction $\zeta$ is the curvature perturbation $\Psi$ in the above defined co-moving gauge i.e $\zeta=\Psi_{|\delta\phi=0}$. It is easy to check that the quantities on the right hand side of \eqref{7.2} can be computed in any gauge. We will perform all our calculations using $\zeta$ and for simplicity we will assume $\frac{\lambda}{3!}\dot{\zeta}^{3}$ interaction(see section \ref{sec:interactions}) to facilitate comparison with the results of \cite{Green:2020whw} throughout. 

To transition to a quantum theory of these perturbations, one needs to define a vacuum state and the accompanying basis of operators. We will largely be concerned with 2 kinds of vacuum states-the Bunch Davies vacuum, and the $\alpha$-vacuum. Perturbation theory in alpha vacuum is subtle, for reasons including but not limited to the difficulties in regularization of the energy momentum tensor. The most common choice for a dS invariant background is the BD prescription for modes, and when we consider excited states, we will introduce them as excited for a Bunch Davies observer. We will then see what happens when there exist observers(related by a Bogolyubov transform) whose vacuum happens to coincide with \textit{some} of these excited states-in particular the $\alpha-$vacuum. We will work with a derivative interaction as it is mathematically simple, and despite being an irrelevant operator that is ideally suppressed by some UV energy scale, it's imprints on nongaussianities is important-in fact, a theory with these interactions shows greater NG enhancement than a theory without them \cite{Holman:2007na}. This is because these interactions describe the IR behaviour of a high energy theory \textit{today}, and the scales accessible at low energies today must have been in the UV in the far past-this is a direct consequence of the fact that in dS irrelevant interactions are accompanied by 2 dimensionful quantities-the cutoff scale and the scale factor $a$ that competes with it. This is the crucial modification to effective operators' importance that inflation provides, not to be found in flat space. As mentioned before, this is also mathematically convenient as this setup avoids possible log divergences in the $\tau\to 0$ limit, and the singularity structure is instead simple poles. The functional dependence of these poles might be different in both these cases, but the points of singularities remain the same(logarithmic $k$ poles with $\zeta^3$ and $1/k$ poles with $\zeta'^3$).

  \section{In-In Formalism}
  \label{in-in}
The predictions of inflation are tested by analysing the features of temperature correlations on the CMB surface, which is the earliest moment in the history of the universe observationally accessible to us. These boundary correlators have information about the dynamics and field content during inflation. More, precisely we are interested in computing equal time correlation functions on the CMB surface unlike flat space QFT where are interested in time ordered correlation functions. The object to compute, say in the interaction picture is
\begin{align}
\label{3.1}
    _{I}\!\bra{\psi(t)}\zeta^{I}(\vec{x_1},t)\zeta^{I}(\vec{x_2},t)\zeta^{I}(\vec{x_3},t)....\zeta^{I}(\vec{x_{n}},t)\ket{\psi(t)}_{I}
\end{align}
where $\ket{\psi(t)}$ is the state of the field configuration at time $t$. To make this object tractable we evolve the state backward in time to the far past ($t\rightarrow-\infty$) where we turn off the interaction using an $i\epsilon$ prescription and the state goes to initial free vacuum state(our choice). We identify all pictures in the far past and therefore the state is also the Heisenberg picture state and hence without any time argument.
\begin{align}
\label{3.2}
    \bra{0}U_{I}^{-1}(t,-\infty)\zeta^{I}(\vec{x_1},t)\zeta^{I}(\vec{x_2},t)\zeta^{I}(\vec{x_3},t)....\zeta^{I}(\vec{x_{n}},t)U_{I}(t,-\infty)\ket{0}
\end{align}
where it can be shown\cite{Weinberg_2005} 
\begin{equation}
\label{in}
    U_{I}(t,-\infty)=Te^{-i\int_{-\infty(1-i\epsilon)}^{t}dt' H_{I}(t')}
\end{equation}
where $H_{I}(t)$ is the interaction hamiltonion in the interaction picture. Note that the above discussion is not special to vacuum state only, It can be generalised to any initial state. \par
If we want to compute \eqref{3.1} in usual flat space QFT assuming that the system started in the vacuum in the far past then nothing except the form of $H_{I}$ will change. Although to simplify things further, in flat space one does an additional manipulation to \eqref{3.2} by assuming that the vacuum is stable. This is achieved by evolving the free vacuum in far past to the far future (turning off interactions in far future also) then
\begin{align}
\label{3.4}
    U_{I}(\infty,-\infty)\ket{0(-\infty)}=\ket{0(\infty)}=e^{iL}\ket{0(-\infty)}
\end{align}
therefore, the two vacua just differ by a phase
\begin{equation}
    e^{iL}=\bra{0(-\infty)}U_{I}(\infty,-\infty)\ket{0(-\infty)}
\end{equation}
Now, assuming a stable vacuum, we evolve the bra state in \eqref{3.1} to the far future (turning off the interactions again) instead of far past and for the right ket we do same as before, we get
\begin{align}
     \bra{0(\infty)}U_{I}(\infty,t)\zeta^{I}(\vec{x_1},t)\zeta^{I}(\vec{x_2},t)\zeta^{I}(\vec{x_3},t)....\zeta^{I}(\vec{x_{n}},t)U_{I}(t,-\infty)\ket{0(-\infty)} \nonumber \\
   = e^{-iL} \bra{0(-\infty)}U_{I}(\infty,t)\zeta^{I}(\vec{x_1},t)\zeta^{I}(\vec{x_2},t)\zeta^{I}(\vec{x_3},t)....\zeta^{I}(\vec{x_{n}},t)U_{I}(t,-\infty)\ket{0(-\infty)}\\
  =  \frac{\bra{0(-\infty)}U_{I}(\infty,t)\zeta^{I}(\vec{x_1},t)\zeta^{I}(\vec{x_2},t)\zeta^{I}(\vec{x_3},t)....\zeta^{I}(\vec{x_{n}},t)U_{I}(t,-\infty)\ket{0(-\infty)}}{\bra{0(-\infty)}U_{I}(\infty,-\infty)\ket{0(-\infty)}}\\
    = \frac{\bra{0}U_{I}(\infty,t)\zeta^{I}(\vec{x_1},t)\zeta^{I}(\vec{x_2},t)\zeta^{I}(\vec{x_3},t)....\zeta^{I}(\vec{x_{n}},t)U_{I}(t,-\infty)\ket{0}}{\bra{0}U_{I}(\infty,-\infty)\ket{0}}
\end{align}
In the last line there is no time argument on the states because we have identified all pictures in the far past. In cosmology the situation is even more simple. Since, we are interested in the correlation functions in the far future (end of inflation/future boundary of de Sitter ) therefore, the question of evolving states to the far future does not even arise and we instead work with \eqref{3.2}.

\section{$\alpha$ modes with a one particle initial state}
\label{sec:alpha1particle}
  For completeness, we repeat the calculation for a one particle state, but this time the one particle state is defined for an $\alpha$ vacuum. We will not explicitly write down the normalization in the denominator; it is always present and serves to regulate the momenta configurations where support for physical poles is achieved. The only difference between the calculations of this section and section(\ref{oneparticleBD}) is the mode functions.
 
  \subsubsection{Two-Point Function}
  At $O(\lambda^{0})$,
  \begin{align}
      _\alpha\!\bra{p}\zeta_{k_1}\zeta_{k_2}\ket{p}_{\alpha} =\frac{\Delta^{2}_{\zeta}|\alpha+\beta|^{2}}{k^{3}}\delta^{3}(\vec{k_1}+\vec{k_2})\left[1+\frac{1}{V}\left(\delta^{3}(\vec{k_1}+\vec{p})+\delta^{3}(\vec{k_1}-\vec{p})\right)\right]
      \end{align}
  \subsubsection{Three-Point Function}
  At $O(\lambda)$, we wish to compute
  \begin{equation}
      _\alpha\!\bra{p}\zeta_{k_1}\zeta_{k_2}\zeta_{k_3}\ket{p}_{\alpha}
        \end{equation}
        Following similar procedure as highlighted in Section 8.2, we arrive at the following result,
%   \begin{eqnarray}
%    \begin{split}
 %       =\frac{-4\lambda\Delta^{6}_{\zeta}H^{-1}(\alpha+\beta)^{3}}{k_1k_2k_3}\delta^{3}(\vec{k_1}+\vec{k_2}+\vec{k_3})\left[\left(\beta^{*}^{2}\alpha^{*}-\alpha^{*}^{2}\beta^{*}\right)\left(\frac{1}{(k_1+k_2-k_3)^{3}}+\frac{1}{(k_1-k_2+k_3)^{3}}\\+\frac{1}{(-k_1+k_2+k_3)^{3}}\right))+\left(\beta^{*}^{3}-\alpha^{*}^{3}\right)\frac{1}{(k_1+k_2+k_3)^{3}}\\+\delta^3({\vec{k_1}-\vec{p}})\left[\left(\beta^{*}^{2}\alpha^{*}-\alpha^{*}^{2}\beta^{*}\right)\left(\frac{1}{(k_1+k_2-k_3)^{3}}+\frac{1}{(k_1-k_2+k_3)^{3}}\\+\frac{1}{(-k_1+k_2+k_3)^{3}}\right))+\left(\beta^{*}^{3}-\alpha^{*}^{3}\right)\frac{1}{(k_1+k_2+k_3)^{3}}\right]\\+\delta^3({\vec{k_1}+\vec{p}})\left[\left(\beta^{*}\alpha^{*}\beta-\alpha^{*}\beta^{*}\alpha\right)\left(\frac{1}{(-k_1+k_2-k_3)^{3}}+\frac{1}{(-k_1-k_2+k_3)^{3}}\\+\frac{1}{(k_1+k_2+k_3)^{3}}\right))+\left(\beta^{*}^{2}\beta-\alpha^{*}^{2}\alpha\right)\frac{1}{(-k_1+k_2+k_3)^{3}}\right]
%        \\+k_1\leftrightarrow k_2+k_1\leftrightarrow k_3\right]
 %        \end{split}
  %  \end{eqnarray}

\begin{eqnarray}
&&=\frac{-4\lambda\Delta^{6}_{\zeta}H^{-1}(\alpha+\beta)^{3}}{2k_1k_2k_3}
\delta^{3}(\vec{k_1}+\vec{k_2}+\vec{k_3}) \nonumber \\
&& \Bigg[ \bigg\{ \left(\beta^{* \, 2}\alpha^{*}-\alpha^{*2}\beta^{*}\right)\left(\frac{1}{(k_1+k_2-k_3)^{3}}+\frac{1}{(k_1-k_2+k_3)^{3}}
+\frac{1}{(-k_1+k_2+k_3)^{3}}\right)  \nonumber \\
&& +\left(\beta^{*3}-\alpha^{*3}\right)\frac{1}{(k_1+k_2+k_3)^{3}} \bigg\}
\nonumber \\
&& +\frac{\delta^3({\vec{k_1}-\vec{p}})}{V} \bigg\{ \left(\beta^{*2}
\alpha^{*}-\alpha^{*2}\beta^{*}\right)\left(\frac{1}{(k_1+k_2-k_3)^{3}}+
\frac{1}{(k_1-k_2+k_3)^{3}} \right. \nonumber \\
&& \left. +\frac{1}{(-k_1+k_2+k_3)^{3}}\right)+
\left(\beta^{*3}-\alpha^{*3}\right)\frac{1}{(k_1+k_2+k_3)^{3}} \bigg\}
\nonumber \\
&& +\frac{\delta^3({\vec{k_1}+\vec{p}})}{V}
\bigg\{ \left(\beta^{*}\alpha^{*}\beta-\alpha^{*}\beta^{*}\alpha\right)
\left(\frac{1}{(-k_1+k_2-k_3)^{3}}+\frac{1}{(-k_1-k_2+k_3)^{3}} \right. \nonumber \\
&& \left. +\frac{1}{(k_1+k_2+k_3)^{3}}\right)+\left(\beta^{*2}\beta-\alpha^{*2}\alpha\right)\frac{1}{(-k_1+k_2+k_3)^{3}}\bigg\} \nonumber \\
&& +k_1\leftrightarrow k_2+k_1\leftrightarrow k_3 \Bigg]+\textit{c.c}
\end{eqnarray}
    
    We see that these states always display a physical pole, and one doesn't require to go to isolated points in phase space to recover these. This is to be expected, because of the nature of mode mixing in $\alpha$ vacuum mode functions. As before, exciting all momenta also removes the isolated nature of some of these physical poles, although it is necessary to do so in this case.

  \section{Review  of Coherent states}\label{sec:review}
  
  Recall that the vacuum state of a harmonic oscillator, being a gaussian, is a minimum uncertainty wavepacket. Starting with
  
  \begin{equation}
      \hat{x}^2=\frac{\hbar}{2m\omega}(a+a^\dagger)^2, \hat{p}^2=-\frac{m\omega\hbar}{2}(a-\a^\dagger)^2
  \end{equation}
  
  And noting that $\langle \hat{x}\rangle=0=\langle \hat{p}\rangle$, we see that 
  
  \begin{equation}
      \langle0| \hat{x}^2|0\rangle\langle0| \hat{p}^2|0\rangle=\frac{\hbar^2}{4}
  \end{equation}
  
  Yet another state with this property is a coherent state $|C\rangle$, defined so that
  
  \begin{equation}
      a|C\rangle=c|C\rangle
  \end{equation}
  
  where $c$ is a complex number. Since $$\langle C|a+a^\dagger|C\rangle=(c+c^*)=2\textrm{Re}(c),\langle C|a-a^\dagger|C\rangle=(c-c^*)=2i\textrm{Im}(c)$$ and $$\langle C|(a+a^\dagger)^2|C\rangle=4(\textrm{Re}(c))^2+1, \langle C|(a-a^\dagger)^2|C\rangle=-4(\textrm{Im}(c))^2-1$$
  
  We find that the uncertainty relation is indeed saturated for these states
  \begin{equation}
      \langle(\Delta \hat{x})^2\rangle\langle(\Delta \hat{p})^2\rangle=\frac{\hbar^2}{4}
  \end{equation}
  
  An exact expression for coherent states is
  
  \begin{equation}
      |C\rangle=e^{-\frac{|c|^2}{2}}\sum_{n=0}^\infty\frac{c^n}{\sqrt{n!}}|n\rangle
  \end{equation}
  
  which can be expressed in terms of a unitary \textit{displacement operator} acting on the vacuum
  
  \begin{equation}
      |C\rangle=D(c)|0\rangle,\hspace{5mm} D(c)=e^{ca^\dagger-a^*a}
  \end{equation}
  
  Interestingly, coherent states are complete but not orthogonal, in fact 
  
  \begin{equation}
      |\langle\alpha|\beta\rangle|^2=e^{-|\alpha-\beta|^2}
  \end{equation}
  
  Importantly, the coherent states remain coherent under time evolution, and we can define a time dependent eigenvalue $c(t)=c(0)e^{-i\omega t}$ for $|C(t)\rangle$. Then, the time evolution of the expectation values is(where $\alpha(0)\equiv |\alpha(0)|e^{i\phi}$ in terms of its norm and phase)
  \begin{equation}
      \langle C(t)|\hat{x}|C(t)\rangle\equiv\langle x(t)\rangle_C=\sqrt{2\hbar/m\omega}|\alpha(0)|\cos(\omega t-\phi)
  \end{equation}
  
  which is the equation obeyed by a \textit{classical} harmonic oscillator. In contrast, note that $\langle  \hat{x}\rangle_{S.H.O}=0$. Thus, coherent state is a superposition state which most closely resembles the classical oscillator. Similarly, one can show
  \begin{equation}
    \langle p(t)\rangle_C=-\sqrt{\frac{2m\hbar}{\omega}}|\alpha(0)|\sin(\omega t-\phi)
  \end{equation}
  
%  \textcolor{blue}{Isn't this always true by ehrenefest's theorem that expectation values for classical EOM(probably not in some cases, but i guess they do)} 
  Note that the phase $\phi$ of $\alpha(t)=|\alpha(0)|e^{i\phi}$ can always be chosen to set $\langle x(t)\rangle_C=0$ for some $t$. For example, we can choose $\phi=\frac{\pi}{2}$ to set $\langle x(0)\rangle_C=0$. This ofcourse doesn't remain intact under time evolution, but we are just choosing an initial value for a second order differential equation.
  
\color{black}
\bibliographystyle{JHEP}
{\footnotesize
\bibliography{biblio}}
\end{document}